\documentclass[12pt,preprint]{aastex}

\newcommand{\s}{s$^{-1}$}
\newcommand{\one}{180$^\circ$}
\newcommand{\FP}{Fabry-P\'{e}rot}

\slugcomment{Draft Version October 31 2008}

\shorttitle{}
\shortauthors{}

\usepackage{graphicx}

\begin{document}


\title{Vector Magnetic Fields and Electric Currents from the Imaging Vector Magnetograph}

\author{Jing Li}
\email{jing@ifa.hawaii.edu}
\affil{Institute for Astronomy, University of Hawaii, 2680 Woodlawn
Drive, Honolulu, HI 96822}
\author{A. A. van Ballegooijen}
\email{vanballe@cfa.harvard.edu}
\affil{Harvard-Smithsonian Center for Astrophysics, 60 Garden
Street. Cambridge, MA 02138}
\and
\author{Don Mickey}
\email{mickey@ifa.hawaii.edu}
\affil{Institute for Astronomy, University of Hawaii, 2680 Woodlawn
Drive, Honolulu, HI 96822}

\begin{abstract}
First, we describe a general procedure to produce high quality vector magnetograms using the Imaging Vector Magnetograph (IVM) at Mees Solar Observatory. Two IVM effects are newly discussed and taken into account: (1) the central wavelength of the \FP~is found to drift with time as a result of undiagnosed thermal or mechanical instabilities in the instrument; (2) the Stokes $V$-sign convention built into the IVM is found to be opposite to the conventional definition used in the study of radiative transfer of polarized radiation. At the spatial resolution $2\arcsec \times 2\arcsec$, the Stokes $Q,U,V$ uncertainty reaches $\sim 1\times 10^{-3}$ - $5\times 10^{-4}$ in time-averaged data over 1-hour in the quiet Sun. When vector magnetic fields are inferred from the time-averaged Stokes spectral images of FeI 6302.5 \AA, the resulting uncertainties are on the order of 10 G for the longitudinal fields ($B_{\parallel}$), 40 G for the transverse field strength ($B_{\bot}$) and $ \sim9^\circ$ for the magnetic azimuth ($\phi$). The magnetic field inversion used in this work is the ``Triplet'' code, which was developed and implemented in the IVM software package by the late Barry J. LaBonte. The inversion code is described in detail in the Appendix.

Second, we solve for the absolute value of the vertical electric current density, $|J_z|$, accounting for the above IVM problems, for two different active regions. One is a single sunspot region (NOAA 10001 observed on 20 June 2002) while the other is a more complex,  quadrupolar region (NOAA10030 observed on 15 July 2002). We use a calculation that does not require disambiguation of {\one} in the transverse field
directions. The $|J_z|$ uncertainty is on the order of $\sim 7.0$ mA m$^{-2}$. The vertical current density increases with increasing vertical magnetic field. The rate of increase is about $1-2$ times as large in the quadrupolar NOAA 10030 region as in the simple NOAA 10001, and it is more spatially variable over NOAA 10030 than over NOAA 10001. 

\end{abstract}


\section{Introduction}

The Imaging Vector Magnetograph (IVM) was built in the early 1990s at the Mees Solar Observatory, University of Hawaii, to measure vector magnetic fields on the Sun.  The IVM is a complicated instrument requiring care in both taking of the data and the subsequent calibration and analysis for the determination of accurate vector magnetic fields.  Characteristics of the instrument and the data calibration have been described in a series of published papers \citep{mickey1996, labonte1999,leka2001,labonte2004}.  The present paper is a continuation of this series in which we identify and discuss two important issues with the IVM data and analysis that were previously not acknowledged in the literature. 

The IVM is designed to combine the advantages of two distinct types of magnetograph, namely, the narrow-band filter polarimeters
and the spectroscopic polarimeters. Both types are based on the Zeeman
effect with the use of magnetic-sensitive spectral lines. They both are in use, for example, onboard the newly
launched HINODE spacecraft \citep{2007SoPh..243....3K}.  

Narrow-band filter magnetographs employ birefringent filters for wavelength selection \citep{1982SoPh...80...33H,ai1986,1995SoPh..159..203Z}. By rotating the crystal elements inside the filter, a narrow band spectral image of a polarization state is formed at a magnetically sensitive
spectral line position. The advantage of the filter devices is that high spatial and temporal resolution 2-dimensional polarization images can be formed at high signal-to-noise ratios. It takes only $\sim$10 minutes to complete a single vector magnetogram. The disadvantage is that the four Stokes parameters are not determined simultaneously within this period, so that spurious polarization may be introduced due to changes on the Sun and in the Earth's atmosphere. Neither is the entire spectral line observed, which limits the diagnostic methods used in the magnetic field inversion process. The magnetic field filling factors cannot be estimated with a narrow-band filter magnetograph \citep{lites1994} and magneto-optical effects may be underestimated in the measurements of transverse fields \citep{2000SoPh..191..309H}.

Spectro-polarimeters record the solar images using a narrow slit placed in front of a spectrograph. Stokes parameters are measured as functions of wavelength across a spectral line sensitive to the Zeeman effect. High spectral resolution can be achieved. To minimize the influence of time-dependent seeing and solar variations, one such spectro-polarimeter, the Advanced Stokes Polarimeter (ASP), uses two CCD cameras to record orthogonal polarization states simultaneously \citep{lites1994,1996SoPh..163..223L}.  The entire Stokes spectral profiles are taken into account for inferring the magnetic fields and the magneto-optical effect and magnetic filling factor can both be well estimated. A disadvantage is that this kind of
instrument takes a longer time, perhaps $\sim$20 minutes to an hour, to obtain a single vector magnetogram  at high signal-to-noise ratios
\citep{1992SPIE.1746...22E,lites1994,1996SoPh..163..223L}. Furthermore, the observing field of view is often smaller than those of narrow band polarimeters. The Haleakala Stokes Polarimeter (HSP) also belongs to this category \citep{mickey1985,1993ApJ...411..362C}.
 
The Mees IVM measures the vector magnetic field in a slightly different way from the above instruments.  A \FP~interferometer is employed to scan the entire magnetically sensitive spectral line (FeI $\lambda6302.5$) at each imaging pixel with a spectral resolution of 0.04 \AA~and a wavelength coverage of 1.2 \AA.   In principle, the IVM is in the spectro-polarimeter category but it has a wide field of view, $4.7\arcmin \times 4.7\arcmin$. The design aims at tackling the atmospheric seeing variation, at the same time acquiring data at high temporal ($\sim2$ minute data cadence) and spatial ($0.55\arcsec \times0.55\arcsec$ pixel size) resolution. The polarization precision is aimed at 0.001 \citep{mickey1996}. 

As pointed out by \citet{labonte1999}, two basic steps are involved in the reduction of the IVM data.  Step 1: deduce the Stokes parameters $I,Q,U,V$ from the raw IVM data. The Stokes $I$ represents the total intensity of the beam; the Stokes $Q,U$ represent states of linear polarization in which the electromagnetic waves are in planes separated by $45^{\circ}$ with planes are perpendicular to the direction of the beam; the Stokes $V$ represents the circular polarization state in which the electromagnetic waves rotate around and propagate along the direction of the beam. The IVM software package was written to deal with many issues in the data in this step  \citep{mickey1996, labonte1999,leka2001,labonte2004}. Step 2: infer the vector magnetic field from Stokes parameters via the radiative transfer equations for polarized spectral light in the presence of magnetic fields. This step is dependent on Stokes inversion methods.  Three methods are implemented in the IVM package: ``wavelet method'', ``derivative method'' and the ``triplet'' method.  A number of investigations based on IVM data used the first two methods [e.g.]\citep{1999SoPh..188....3L,2002A&A...392.1119R,2002A&A...395..685B,2003ApJ...595.1277L,2003ApJ...595.1296L,2004ApJ...610.1148W,2005ApJ...623L..53M,2005ApJ...629..561B,leka2007}.  The current work will use the ``triplet'' method which will also be described in detail in the Appendix A. The ``triplet'' method was also used in the studies based on IVM data by a few [e.g.]\citep{2004ApJ...615.1029G, 2006A&A...451..319R,2006ApJ...636..475G,2007ApJ...671.1034G}. 

Recently, two problems in the IVM data caught our attention.  Neither of them has been discussed in the literature, nor have they been taken into account in the standard IVM package. The effects of these two problems are examined in this paper. The first problem is that the spectral line center produced by the \FP~shifts with time in the raw data. The cause of this problem is unknown, but is presumably related to a mechanical/thermal shift in the \FP~causing the spectral line to drift throughout the day.  If uncorrected, the wavelength shift in 1 hr reaches about 0.2 \AA, equivalent to the (very significant) Doppler shift of $9.5$ km s$^{-1}$.   We will discuss a procedure for minimizing the error caused by the shift in the Stokes data without jeopardizing the solar velocity measurements.  We do not know for how long the spectral line drift has been a problem with IVM: conceivably it affects all data taken with this instrument.

The second problem is that the sign convention built into the IVM determination of  Stokes $V$ is unfortunately opposite to the conventional definition employed in the analysis of the radiative transfer of polarized radiation. A brief check shows that the problem was not present in the data taken on 19 June 1998, but was present in data taken on 15 July 2002, and on 6 August 2003 (see Figure 5 of \citet{labonte2004}).  According to the (anonymous) referee of this paper, the problem has been unofficially discussed among IVM users, notably the late Tom Metcalf, who mentioned the Stokes $V$-sign problem. By all the evidence, we suggest that the Stokes-$V$ signs were correct before January 1999 with image size $256\times 256$. The signs are wrong for the data taken after January 1999 when the IVM underwent a major upgrade, and acquired a larger CCD camera allowing image size $512\times 512$.  The Stokes $V$-sign problem results in over-correction of the magneto-optical effect in the Stokes polarization measurements and must be removed if accurate inversions are to be obtained.

An important application of vector magnetograph is to determine the vertical component of the electric current density in the photosphere, which is an important measure of the non-potentiality of the active regions, the helicity content. It has the potential to illuminate the active region eruption process.  Electric current densities have been repeatedly obtained since the photospheric vector magnetic fields became available in 1980s [e.g.]\citep{1984SoPh...91..235D,1987SoPh..109...81L,1988SoPh..115..107H,1992PASJ...44L.111C,1993ApJ...411..370L,1993ApJ...411..378D,1994SoPh..151..129Z,1994SoPh..155...99W,1994SoPh..149..309V,1994ApJ...428..860M,1995ApJ...445..982G,1997ApJ...482..490L,2004ApJ...606..565B,2004ApJ...615.1029G,2008AdSpR..42..888G}. All these calculations involve the resolution of the \one~ambiguity in the transverse field directions \citep{2006SoPh..237..267M}. In this paper, we will calculate the absolute vertical electric current density, $|J_{z}|$, with the equation derived by \citet{1998A&A...331..383S} which does not require the \one~disambiguation of the  transverse field directions. Therefore, this removes one uncertainty in the current calculation. The $|J_{z}|$ is particularly useful for studying the mechanical forces due to currents induced in moving material \citep{1945RSPSA.183..453C}.

In section 2, we give a brief description of the IVM instrument and the data reduction principles. Sections 3 and 4 correspond to the steps 1 and in the IVM data reduction. We describe the procedure used to generate high-quality Stokes images from IVM raw data with reference to the active region NOAA 10001 on 20 June 2002 in section 3, and  describe the procedure to generate the vector magnetic fields from Stokes images in section 4. The absolute vertical current densities, $|J_z|$, are presented in section 5 for two very different active regions, NOAA 10001 of 20 June 2002, and 10030 of 15 July 2002. Section 6 gives a summary of the results. A detailed description of the magnetic field inversion code, ``Triplet Fitter'', is given in the Appendix.

\section{IVM Data Reduction}

The IVM has a square field-of-view 4.7 arcmin on a side. The basic components of the instrument include a polarization modulator, a tunable \FP~filter, a beamsplitter, pre-filters, and Charged-Couple Device (CCD) cameras \citep{mickey1996}. Among the system components, the \FP~ is
the fastest at scanning the spectral profiles. The time resolution of the system is limited by the CCD camera readout. For the photospheric
magnetic field observations, the \FP~filter is centered at FeI $\lambda6302.5$, with a wavelength coverage of 1.2 \AA~and a spectral resolution of 0.04 \AA.  Currently, the IVM detector is a $512\times 512$ pixel CCD array. Each pixel corresponds to $0.55\arcsec\times 0.55\arcsec$ spatial size and the readout rate per pixel is 320 kHz. It takes 0.8 second to read out the whole array, in which time interval is made to a change among
the four polarization states, and a scan of a new wavelength position. Therefore, an IVM data set including 30 wavelength positions in four polarization states can be completed in $\sim 1.6$ minutes.  

A special feature of the IVM is that two practically identical cameras record the image simultaneously: the so-called ``Data'' and ``Geometry'' cameras. Their purpose is to minimize the ``seeing'' effect on the polarization signals obtained in the 4 polarization states at different times. The only significant difference between the two cameras is the filters in front of them: a narrow band 3 \AA~filter for the ``Data'' camera to select a single order from the \FP~centered at the FeI $\lambda6302.5$, and a broad band 100 \AA~filter for the ``Geometry'' camera to select roughly 30 orders from the \FP. The idea is that the ``Data'' camera records the polarization images, and the ``Geometry'' camera records the (nearly) continuum
images. Any brightness fluctuations in the ``Geometry'' camera must be due to ``seeing'' variations, while changes recorded by the ``Data'' camera are due to polarization signals from the Sun. 

A standard IVM calibration package has been implemented in IDL procedures based on the IVM observational model described by
\citet{labonte1999}. In the package, the correction of the dark current, flat field, scattered light and polarization cross-talk are first executed. The effects of the ``seeing'' on the images include blurring, displacing and distorting the fine solar structures. The best ``Geometry'' image is chosen within a single data set as a reference image. Shifting, de-blurring and de-stretching functions are then applied to the whole set of ``Geometry'' images. These very same corrections are then applied to the set of ``Data'' images \citep{mickey1996,labonte2004}.  The final output consists of the calibrated Stokes images, $I,Q,U,V$. 
 
\section{Step 1: Stokes Parameters Reduction }

\begin{figure}[]
\begin{center}
\includegraphics[width=1.0\textwidth]{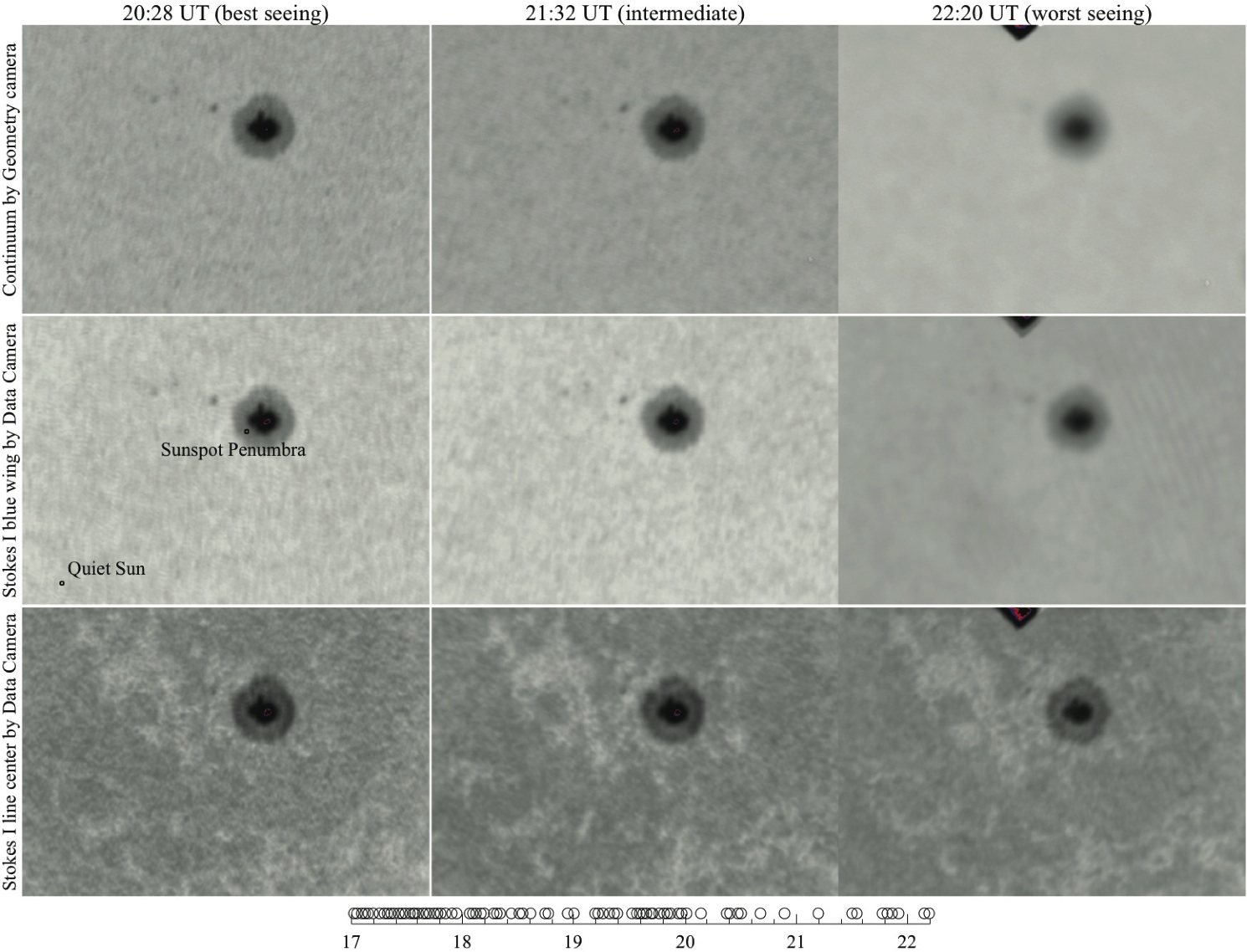}
\caption{Active region NOAA 10001 on 20 June 2002 (N20W07). From top to bottom, images are the photospheric continuum taken with the ``Geometry'' camera (top row), Stokes $I$ at the blue wing (middle row) and the Stokes $I$ at the spectral line center (bottom row) by the ``Data'' camera. From left to right, images represent the best seeing, $\epsilon=0.0062$ at 20:28 UT, the intermediate seeing, $\epsilon=0.0038$ at 21:32 UT and the worst seeing, $\epsilon=0.0013$ at 22:18 UT. The circles on left-middle image mark the ``Quiet Sun'' and ``Sunspot Penumbra'' have the radius, $r=2$ pixels equivalent to $2\arcsec \times 2\arcsec$. All images have the same size $240\arcsec \times 170\arcsec$. The plot with circles on the bottom of the figure shows the selected IVM data sets as function of time [hours] based on the seeing criterion with seeing condition $\epsilon \geq 0.0038$. \label{fig1}}
\end{center} 
\end{figure}

In this and section 4, we describe the reduction procedures 
with reference to the active region NOAA 10001. In this section, we describe the Stokes parameters reduction corresponding to the IVM data reduction step 1. On 20 June 2002 this AR consisted of a single
spot of area 200 [$10^{-6}$ of solar hemisphere] surrounded by a plage region and projected near the
disk center N20W07.  It was observed continually from 17:00 to 22:20 UT
by the IVM, giving a total of 149 vector magnetograms. Fig.\ref{fig1} shows AR10001 on 20 June
2002 taken in different seeing conditions in the continuum by the ``Geometry'' camera (top row), and the Stokes $I$ by the ``Data'' camera at the blue wing (middle row) and the center of the spectral line FeI 6302.5 \AA. 

\subsection{Data Selection based on the Seeing Condition}

Although it has been best-corrected by the ``Geometry'' images, the remaining uncorrected ``seeing''  is the major source of
the noise in the calibrated Stokes images. \citet{leka2001} implemented an algorithm for seeing estimation into the IVM software package, which is based on the concept of the Modulation Transfer Function.  For each ``Geometry'' image, $G$, the
quantity, $\epsilon$, is calculated from root-mean-square ({\it r.m.s.}) 

\begin{equation}
\epsilon(\lambda,t) = r.m.s. \{[G(\lambda,t)-G_s(\lambda,t)]/G_s(\lambda,t)\}
\label{epsilon}
\end{equation}


\noindent where $G_s$ is the image smoothed with a boxcar average. $\epsilon(\lambda,t)$, as a function of wavelength $\lambda$ at time $t$, in fact, measures the image contrast. \citet{leka2001} commented ``When this seeing measure
decreases towards 0.005, granulation is no longer visible and the
seeing is estimated 2 arc sec.''  Ideally, $\epsilon(\lambda,t)$ should be invariant with the polarization state. In reality, the ``Geometry'' camera receives the same beam as the ``Data'' camera which is modulated by the polarization modulator. As a result, the Geometry images include the polarization characteristics of the spectral line but diluted by a factor of $1/30$ because one of 30 \FP~orders falls in the line FeI 6302.5 \AA. To judge the seeing condition with continuum images, we use the values of $\epsilon(\lambda,t)$ at the most-blue wing.  

For the 20 June 2002 data, about 20\%
of the data satisfy $\epsilon \geq0.005$. Since we are mostly interested in the sunspot and its penumbra, we relax the seeing threshold to
$\epsilon=0.0038$ (see the middle column images in Fig. \ref{fig1}) which allows 53\% of the total 149 data sets
to be selected for further processing. The time distribution of these data sets, which will be used
to generate the vector magnetograms, is indicated with
circles along the horizontal axis in the bottom plot of Fig.\ref{fig1}. The images from the left to right are IVM images with the best seeing ($\epsilon=0.0062$), the seeing value used as the threshold for the current
data selection ($\epsilon=0.0038$), and the worst seeing ($\epsilon=0.0013$). On average, the seeing is $\sim 2$\arcsec, which will be the minimum pixel sample size in the following data analysis.

\subsection{Spectral Line Shift}

One of the system imperfections is that the central wavelength
varies with spatial position on the detector. The central wavelength varies quadratically with position in
the field because the \FP~is used in the collimated beam. This was discussed by
\citet{labonte1999} and is implemented in the calibration of the flat field in the IVM software
package.  The spurious intensity pattern was ``simply the convolution of the line
profile and the etalon bandpass.'' The quadratic component of the vignetting is measured in the flat-field images taken in the
continuum wavelengths. The flat-field image at each wavelength is
divided by this wavelength-insensitive vignetting. Each flat field
image is fitted again with a quadratic spatial distribution which is
the wavelength-sensitive quadratic fitting. After dividing out the
newly fitted wavelength-sensitive quadratic fitting, the flat-field
images are multiplied by the continuum quadratic vignetting.
 
\begin{figure}[]
\begin{center}
\includegraphics[width=1.0\textwidth]{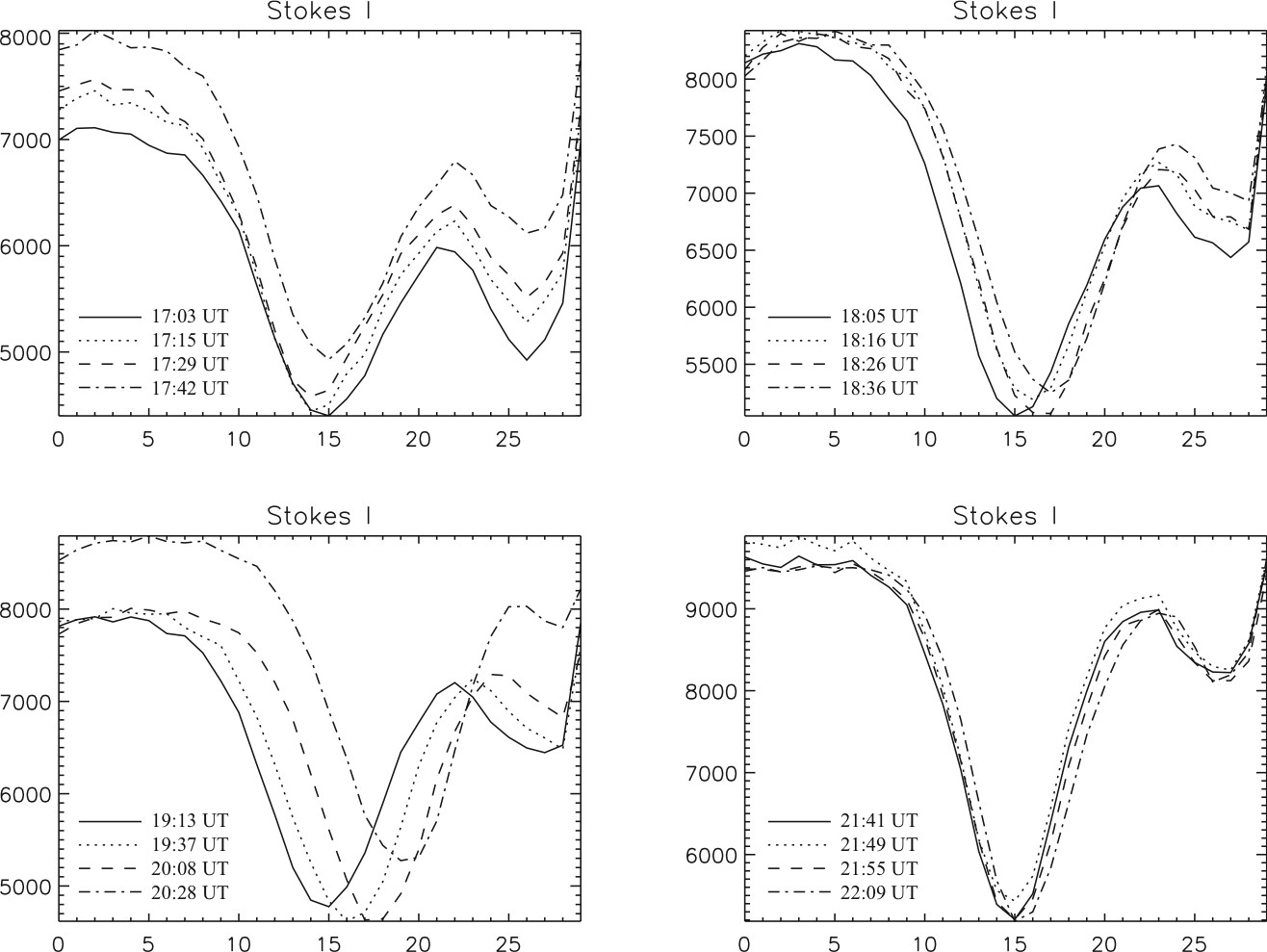}
\caption{Examples of the Stokes $I$ spectral line profiles throughout the day. The profiles are the signals, {\it r.m.s.}[$S_D(t,x,y,I,\lambda)$] over the  circled ``Quiet Sun'' region in Fig.\ref{fig1}. The wavelength shift with time is not uniform in the course of the day. The dips at the wavelength position   $\sim 28$ are due to the telluric $O_2$ line. 
\label{spectrallines}}
\end{center} 
\end{figure}

\begin{figure}[]
\begin{center}
\includegraphics[width=0.85\textwidth]{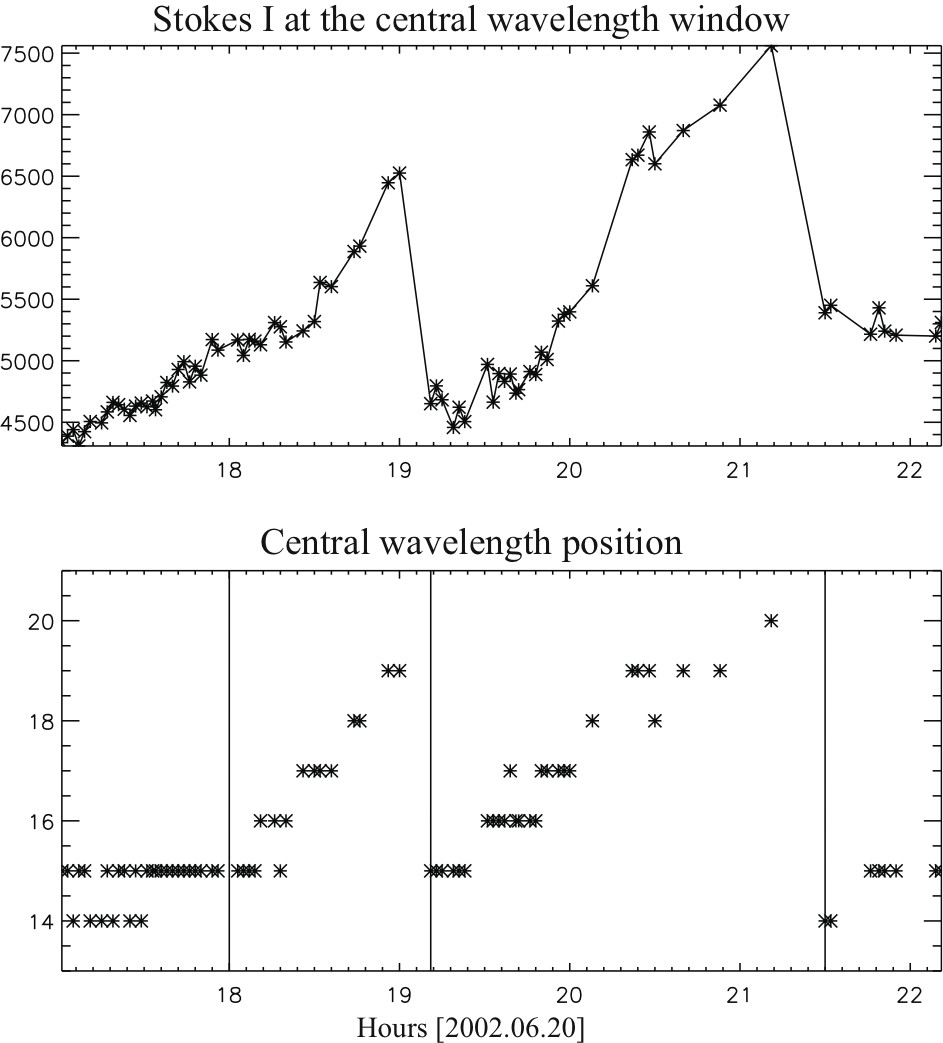}
\caption{Top panel: the Stokes $I$ signals at the central position of the IVM wavelength window as a function of time. The signals are taken over the ``Quiet Sun'' region shown with a circle in Fig.\ref{fig1}. ``$\ast$'' symbols represent the IVM data sets selected by the seeing threshold. Bottom panel: the wavelength positions of the minimum Stokes I intensity as a function of time. The ``$\ast$'' symbols are the same data points as those on the top panel. The solid vertical lines separate the IVM data into 4 groups. They define the data into four groups within which the wavelength shifts were either very small or to the red wing. At $\sim$ 19:12 UT and 21:30 UT, the central IVM wavelength positions were adjusted back to align with the spectral line center by the observer.
\label{stokesI-time}} 
\end{center} 
\end{figure}

A time-dependent drift in the central wavelength is also present, but was not discussed before. This kind of shift is
demonstrated by the Stokes $I$ profiles in Figure
\ref{spectrallines}. For the observations with the ``Data Camera'', we define $S_D(t,x,y,s,\lambda)$ as the Stokes
polarization signals ($s$) at pixel position ($x,y$) as a function of  time ($t$)
and wavelength ($\lambda$), where $s$ represents the Stokes parameters $I,
Q,U,V$. Although the IVM CCD pixel size is $0.55\arcsec \times 0.55\arcsec$, the actual spatial resolution is lower because of the atmospheric disturbance and residual effect of the atmospheric and instrumental ``seeing''.  The Stokes signals discussed here are the average signals over a cluster of pixels. For the current data set, the actual spatial resolution is a circular region with $r=2$ pixels equivalent to $2\arcsec \times2\arcsec$. In Figure \ref{spectrallines}, the Stokes $I$ is calculated as the {\it r.m.s.}[$S_D(t,x,y,s,\lambda)$] over the circular ``Quiet Sun'' region shown in Figure \ref{fig1}.  After observing for an hour, the amount of
the systematic wavelength shift is $\sim 0.2$ \AA~(5 wavelength positions) and, if uncorrected, the spectral line center could move out
of the spectral passband in only $\sim$3 hours. The wavelength drift is probably
caused by the \FP's transparent plate, although in detail we do not know why, and we do not know for how long the
wavelength drift has been a feature of the IVM.  In the data analyzed here, the etalon was periodically re-adjusted (using the method of \citet{mickey2004}) during the observing sequence so that the center of the spectral line re-aligned
with the center of the \FP.   This operation is reflected in
Fig. \ref{stokesI-time} in which the Stokes $I$ signals at the central position of the IVM spectral window increase with time (top
panel) until, at UT 19:00 and 21:20, they jump back as the central wavelength position was manually adjusted. 

In coincidence, the center of the spectral line varied within the IVM spectral window with time (lower panel). We determine the spectral line centers with the minimum Stokes $I$ intensity over the ``Quiet Sun''.  Sometimes, the central positions are one pixel apart in consecutive IVM data sets due to the line center fluctuation. This occurred more frequently when the telluric line was strong in the morning such as before 18 UT in the current data (see Fig. \ref{spectrallines}). 


The complication caused by the spectral line shift is that it is a mixture
of real solar vertical motions due to the Doppler effect and the instrumental
IVM spectral shifts. The uncorrected wavelength drift of $\sim 0.2$ \AA~in one hour means that,
without correction, the IVM line center drifts by as much as 9 km \s. The normal IVM integration is 2 minutes, in which time the line center drifts by about 0.3 km \s. The latter sets an absolute lower limit to the velocity resolution in the data. It appears that this kind of wavelength shift does not vary with the spatial position. 

\subsection{Stokes Parameter Uncertainties}
\begin{figure}[]
\begin{center}
\includegraphics[width=0.8\textwidth]{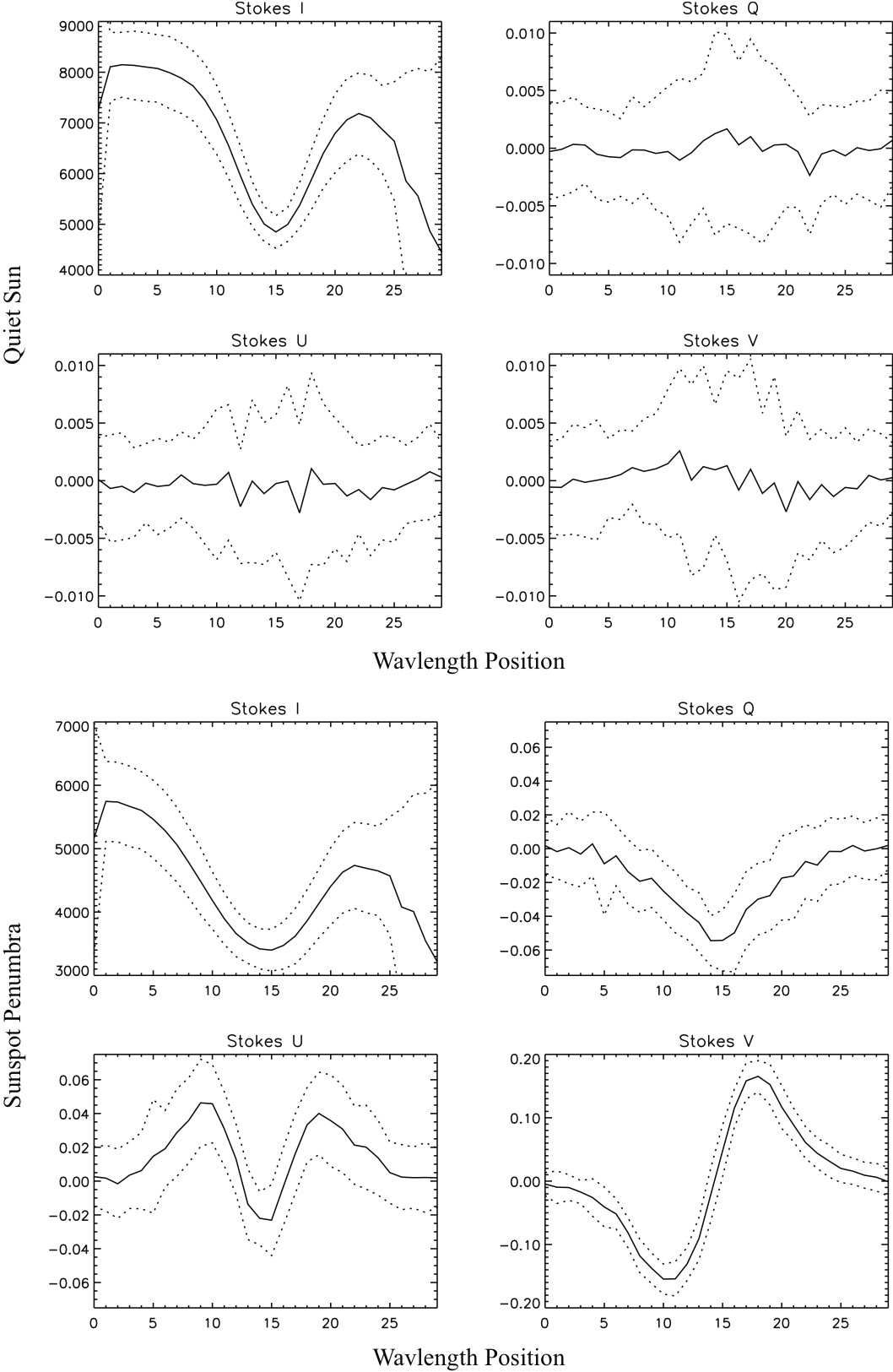}
\caption{The average Stokes signals and the uncertainties measured at two locations: ``Quiet Sun'' (top four panels); and ``Sunspot Penumbra'' (bottom four panels) as function of wavelength. Solid lines: the time-averaging Stokes signals, $\bar S_D(x,y,s,\lambda)$, calculated by averaging IVM ``good seeing'' data sets at each wavelength position; Dotted lines: $\bar S_D(x,y,s,\lambda) \pm \sigma(x,y,s,\lambda)$, where $\sigma$ is the standard deviation of all ``good seeing'' data sets.  \label{stokes_noise}} 
\end{center} 
\end{figure}

Using the ``good seeing'' data set, the time-averaged Stokes signals, $\bar S_D(x,y,s,\lambda)$, and the standard deviations, $\sigma(x,y,s,\lambda)$, represent the statistical averages and uncertainties of the single measurement. In Figure \ref{stokes_noise}, $\bar S_D(s,x,y,\lambda)$ (solid lines) and $\bar S_D(s,x,y,\lambda) \pm \sigma(s,x,y,\lambda)$ (dotted lines) are plotted from
two locations, ``Quiet Sun'', and ``Sunspot Penumbra'' (see circled locations on the middle panel of the first column in the Fig.\ref{fig1}). 
On average, the Stokes $Q,U,V$ had uncertainties $\sigma= 0.005$ in the ``Quiet Sun'', and $\sigma=0.02$ in the
``Sunspot Penumbra''. The signal-to-noise ratios are about 2 for $Q,U$ and 7 for $V$ in both regions. 

The Stokes uncertainties can be reduced by averaging time series
IVM data sets. In the bottom panel of Figure \ref{stokesI-time}, the IVM data sets are divided into
four groups by vertical lines. Each group is defined by both the ``good seeing'' condition and the times at which the \FP~was manually adjusted: 17:01-17:54 UT (27 data sets), 18:03-18:56 UT (15 data sets), 19:00-21:20 UT (28 data sets), and 21:30-22:11 UT (8 data
sets). Within each group, Stokes spectral images are averaged at each pixel and wavelength position after (1) spatially registering each image to the first one; (2) spectrally registering the spectral line to the central IVM spectral window based on the measurements in the ``Quiet Sun'' (see the bottom plot in the Fig. \ref{stokesI-time}). The signal-to-noise ratios are increased by $\sim \sqrt{N}$, where $N=27,15,28,8$, for each averaged Stokes image,  respectively. As a result, the signal-to-noise levels are enhanced to 10 maximum for Stokes $Q,U$, and 37 for Stokes $V$. The Stokes uncertainties are 1-2 $\times 10^{-3}$ in the quiet Sun regions. The four time-averaged Stokes images will be used as input to the ``Triplet'' inversion code for generating the vector magnetograms. 

The time-series averaged Stokes images are highly recommended. Averaging (1) reduces the Stokes polarization uncertainties, and increases the signal-to-noise ratios; (2)  smoothes out rapid fluctuations between the polarization modulations. [These fluctuations are probably real on the Sun, but are impossible to be distinguished by the IVM modulation mode]; (3) eliminate oscillatory effects such as the 5 minute oscillation  \citep{2006ApJ...636..475G}.

\section{Step 2: Photospheric Vector Magnetic Fields}
This section corresponds to the IVM data reduction step 2. The vector magnetic fields in NOAA 10001 are inferred from the time-averaged Stokes
parameters using the ``Triplet'' code. The latter was developed and implemented in the IVM software package by the late Barry
J. LaBonte. Because the code has not been described in any published literature, a detailed description is given in
Appendix A of this paper. In this section, we will briefly describe the magnetic field inversion methods in the IVM package. The final vector magnetograms will be presented in the second part. The third and fourth sections will demonstrate the effects on the magnetograms due to the spectral line shifts with time, and the wrong Stokes $V$-sign set by the IVM hardware.  

\subsection{Magnetic Inversion Codes With IVM}

The general problem is to solve the radiative
transfer equations for polarized light in the solar
atmosphere. Many pioneering works have been put
forward to lay out the foundation for the solutions
\citep{unno1956,auer1977,landolfi1982,skumanich1987,jls1989,rees1989}. The
IVM data reduction package adopts ultimately three methods: (1) the
wavelet method; (2) the derivative method; and (3) the ``Triplet
Fitter''. Here, we briefly describe these three inversion methods. [For the original, but somewhat out-of-date description of
the magnetic field inversion methods in the IVM package, readers are
referred the web site:

http://www.solar.ifa.hawaii.edu/Reference/IVM/IVM\_data\_red.html.]

The wavelet method, developed by the third author, uses the wavelet
transform on each profile to locate position and amplitude of the
Stokes components. The Paul wavelet is used since its real component
is similar to $Q, U$ profiles and its imaginary part matches the $V$
profile. The line center position is computed from the transform of
the $I$ profile, then the $Q,U,V$ amplitudes are obtained from their
transforms at preselected scales. The magnetic field values are
obtained by multiplying the polarization parameters by a ``magic number''.
Thus, the method suffers from both arbitrariness and likely saturation at
large field strengths. The magneto-optical effects are not corrected, neither
are the magnetic filling factors solved. However, this method is quick and,
therefore, has been used, for example,  to reduce a large amount of IVM data for the
statistical study of over 1000 flare-productive active regions
\citep{leka2007} . 

The ``derivative method'' is based on the solution
of radiation transfer equations in the ``weak field'' limit, in which the
Zeeman splitting is a fraction of the Doppler width \citep{jls1989}
(so-called JLS method). Under the weak field condition, Stokes $V$ is proportional to the spectral line slope,
$\partial I / \partial \lambda$, and to the line-of-sight field
$B_\parallel$. The Stokes $Q$ and $U$ are proportional to the
derivative of the spectral line slope, $\partial^2 I / \partial
\lambda^2$, and to the square of the transverse field. With a Doppler width for the
photospheric line of 40 m\AA, the maximum line-of-sight field ($B_\parallel$)
strength measured with FeI 6302.5 (Land\'{e} factor $g_J=2.5$) is
about $\sim$850 G, where Zeeman splitting $4.67\times 10^{-13} g_J
\lambda^2 B_\parallel$ \AA. The method saturates in strong field regions and,
therefore, is unsuited to magnetic field measurements in
sunspots. In order to measure strong magnetic fields, an
alternative method is needed.

The ``Triplet'' code was, therefore, developed as an independent magnetic field diagnosis in the IVM package. 
It originates with the \citet{unno1956} solutions applicable to homogeneous
magnetic fields and to absorption coefficients invariant with respect to optical
depth. The \citet{unno1956} solutions are incorporated into a
non-linear least square method by \citet{auer1977} to infer the
magnetic field from the observed Stokes profiles. The ``Triplet'' code
is based on fitting to a more accurate radiative transfer model of the
line profiles developed by \citet{landolfi1982} who include the
magneto-optical effect to the non-linear least square method. Another special feature of the ``Triplet'' code is that it treats strong and weak magnetic field regions in two separate steps. In the first phase, the photospheric thermodynamic parameters are determined from the continuum measurements in the Stokes $I$ spectra. The magnetic parameters are calculated by fitting
the observed Stokes profiles with equation (\ref{eq:QUVm1}), which
neglects the effects of Faraday rotation. In the second phase, the magnetic fields are re-determined in the sunspot regions only, which are identified by the relative low brightness in the continuum. The
code uses non-linear least square fitting of the observed Stokes
profiles, including Faraday rotation effects, and initial magnetic filling factor $<1$.  The initial values of the model parameters are those
determined from the first phase.  Most thermodynamic parameters remain unchanged, but some of them are allowed to vary.  The basis of two-step treatments is demonstrated in the magnetic field observations using two magnetically sensitive infrared lines \citep{1995ApJ...446..421L}. By fitting the observed Stokes $V$ profile of a normal Zeeman triplet, they found that the photospheric magnetic fields consists of two distinct components: the weak field (typically 500 G) is found in the intranetwork magnetic elements; the strong field (typically 1400 G) is found in the sunspots.  \citet{rmo1987} compared the derivative method
and the least squares fitting \citep{skumanich1987} of the
Unno-Rachkovsky profiles. They concluded that the latter is the
superior technique for deriving the magnetic field parameters. In this
paper, we infer the magnetic fields using the ``Triplet'' code. Before the current paper, the code was used to infer the magnetic fields from IVM data by other authors such as \citet{2004ApJ...615.1029G,2006A&A...451..319R,2006ApJ...636..475G,2007ApJ...671.1034G}.

\subsection{Vector Magnetograms and Uncertainties}
\begin{figure}[]
\begin{center}
\includegraphics[width=1.0\textwidth]{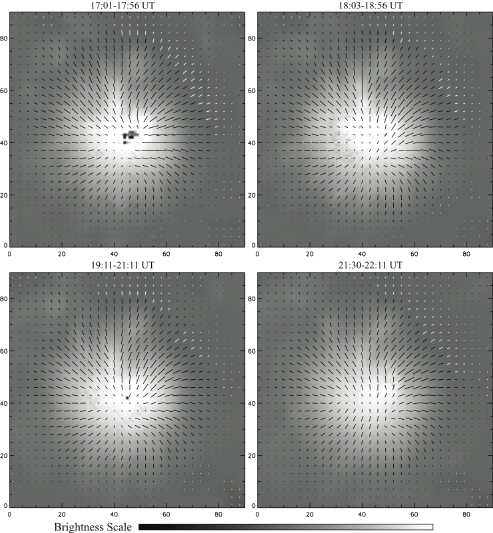}
\caption{NOAA 10001 vector magnetograms derived from time-averaged Stokes images over four time periods: 17:01-17:56 UT, 18:03-18:56 UT, 19:11-21:11 UT, and 21:30-22:11 UT.  The longitudinal fields are the background images, which strengths are shown in the brightness scale [-2500,2500] G. The transverse fields are represented by short bars: white bars are against the negative longitudinal polarity, and the white bars against the positive polarities. The $B_{\bot}$ is represented by bar lengths. The distance between neighboring pixels represents $B\bot=1500$ G. Pixels having $B_\bot < 40$ G are not plotted. The magnetic azimuths are represented by bar orientations. The image sizes are $50\arcsec\times 50\arcsec$. \label{imgms}} 
\end{center} 
\end{figure}

The vector magnetograms are derived from time-averaging Stokes images over
4 time intervals: 17:01-17:56 UT, 18:03-18:56 UT, 19:11-21:11 UT, and 21:30-22:11 UT (Figure \ref{imgms}). Before running the ``Triplet'' code, Stokes-$V$ images are flipped in sign. The issue will be described in detail in section 4.3. To clearly show the transverse field and the magnetic azimuth, the magnetic fields are plotted in size $50\arcsec\times 50\arcsec$ around the major sunspot: the longitudinal fields are the background images;  the transverse fields are presented by short bars, white bars are against the negative polarities, and black bars are against the positive polarities; magnetic azimuths are indicated by the bar directions which remain {\one} ambiguous.

The uncertainties of both longitudinal ($B_\parallel$) and transverse ($B_\bot$) fields are estimated by calculating the {\it r.m.s.}($B_\parallel$), {\it r.m.s.}($B_\bot$) over pixels having the Stokes $Q,U,V$ signals less than $0.001$ representing the quiet Sun. The uncertainties differ slightly among the four final vector magnetograms. The errors are 10 G for the longitudinal magnetic component, 40 G for the transverse magnetic component, and $0.4$ km s$^{-1}$ for the Doppler velocity. Note that the velocity uncertainty is very similar to the spectral line drift within 2 minutes during the IVM standard integration time.

\begin{figure}[]
\begin{center}
\includegraphics[width=1.0\textwidth]{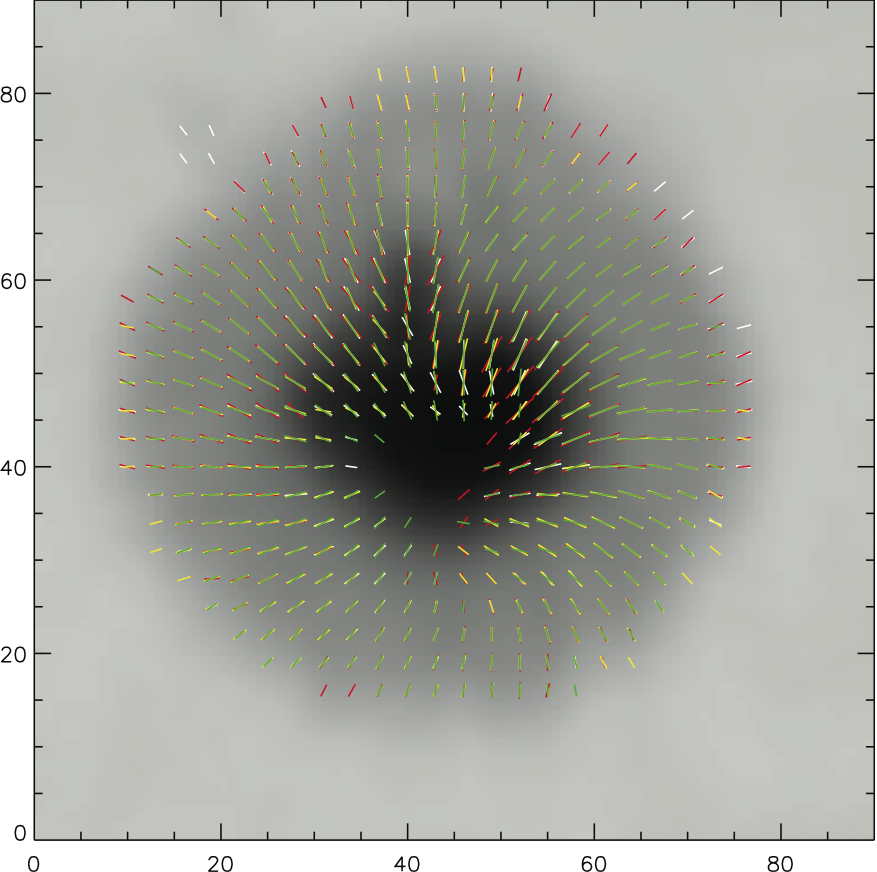}
\caption{The transverse fields ($B_{\bot}>500$ G) are presented by short bars over-plotting the sunspot continuum image. $\bf B_{\bot}$ of four magnetograms are distinguished by bar colors: 17:01-17:56 UT (white); 18:03-18:56 UT (red); 19:11-21:11 UT (yellow) and 21:30-22:11 UT (green). When the bar length is equal to the distance between neighboring pixels, $B_{\bot}=1500$ G. The bar orientations represent the magnetic azimuths. $r.m.s.$ azimuth differences is in average $\sim 9^{\circ}$. The image sizes are $50\arcsec\times 50\arcsec$.  \label{bxys}} 
\end{center} 
\end{figure}

The uncertainties for the magnetic azimuths, $\phi$, are estimated as the {\it r.m.s} azimuth differences between the time-neighboring magnetograms. The differences are functions of the transverse field strengths. The smaller  are the transverse field strengths included, the larger are the magnetic azimuth uncertainties. When the {\it r.m.s} azimuth differences are calculated over pixels having $40<B_{\bot}<500$ G (representing the plage and sunspot edges), the average azimuth uncertainty is $\sim 18.5^\circ$. When the azimuth uncertainty is estimated over the sunspot penumbra ($500<B_\bot<1500$ G), it is on average $\sim 8.9^\circ$. Figure \ref{bxys} shows the transverse fields of the four magnetograms overlapping one another in the sunspot area.

\subsection{Effects of Spectral Line Shifts}
To  quantitatively investigate effects of spectral line shift with time, the four magnetograms described in section 4.2 were used to define the ``correct'' reference magnetograms.  The corresponding time-averaged Stokes images described in section 3.3 were used as images of ``standard Stokes spectra''. All the calculations are conducted around the major sunspot area size 50\arcsec$\times$50\arcsec, as shown in figure \ref{imgms}.

We firstly examine differences between the ``correct'' magnetograms and those derived from time-averaging Stokes images without performing the registration of the spectral line centers. This was the scenario by \citet{2006ApJ...636..475G} who used time-averaging Stokes images over time intervals $20\sim30$ minutes, and \citet{2006A&A...451..319R} who used the time-averaging Stokes images over time interval 15 minutes. The {\it r.m.s.}[$\Delta B_\parallel, \Delta B_\bot, \Delta \phi,\Delta v$] between the two are listed in Table \ref{unc} under ``Experiment 1''.  

For cases of using non-time averaging Stokes images, i.e., single IVM data set, we examine differences between the ``correct'' magnetograms and those  derived from wavelength-shifted Stokes images. This was the scenario by {\it e.g.} \citet{2003ApJ...595.1277L} and \citet{leka2007} who used Stokes images over time intervals of 2 or 4 minutes. The wavelength shifts are simulated by removing a number of pixels in the red wing from the ``standard Stokes spectra'' images (note that these Stokes images have been time-averaged, therefore, the situation is yet different from the real IVM single data set in term of signal-to-noise). The removal is equivalent to missing the part of the spectral red wing, but reader should note that the number of spectral samples is reduced. The comparisons are listed in the Table \ref{unc} under ``Experiment 2'' when the spectral line was shifted by 5 wavelength positions ($\sim 0.2$ \AA) to the red wing. This is normally the maximum wavelength shift before the observer intervenes to adjust the \FP.

The experiments show that the temporal spectral line shifts do not generally affect the magnetic azimuth, neither do they much affect the velocity measurements, but they do affect the measurements of the magnetic field strengths. On the other hand, significant departures in the magnetic field strengths, magnetic azimuth, and the velocity field from the  ``correct'' magnetic field occur when $\sim 1/3$ line profile has drifted out of the IVM spectral window. At this stage, the spectral line center is shifted by $\sim 0.44$ \AA~to the red wing from the center, i.e., leaving IVM running for $\sim 2$ hours without observer's intervention. On average, the {\it r.m.s.}$\Delta B_\parallel \sim1150$, {\it r.m.s.}$\Delta B_\bot\sim 1024$ G, {\it r.m.s.}$\Delta \phi\sim 14^\circ$, and {\it r.m.s.}$\Delta v\sim 18$ km s$^{-1}$.

\begin{deluxetable}{lllll}
\tabletypesize{\scriptsize}
\tablenum{1}
\tablecolumns{5}
\tablecaption{Comparison between the ``Correct'' and ``Incorrect'' Magnetic Fields \label{unc}}
\tablewidth{0pt}
\tablehead{
\colhead{Time [UT]} &
\colhead{$\Delta B_\parallel$ [G]} &
\colhead{$\Delta B_\bot$ [G]} &
\colhead{$\Delta \phi$} &
\colhead{$\Delta v$ [km s$^{-1}$]} 
}
\startdata
\cutinhead{Experiment 1}
17:01-17:56 & 70.5 & 96.5 & $5.9^\circ$ & 0.4 \\
18:03-18:56 & 87.0 & 348.4 & $13.4^\circ$ & 0.2 \\
19:11-21:11 & 48.4 & 96.4 & $6.2^\circ$ & 0.2 \\
21:30-22:11 & 25.4 & 84.9 & $9.4^\circ$ & 0.1\\
\cutinhead{Experiment 2}
17:01-17:56 & 104.9 & 84.7 & $4.6^\circ$ & 0.5 \\
18:03-18:56 & 98.3 & 125.4 & $7.2^\circ$ & 0.2 \\
19:11-21:11 & 36.4 & 76.7 & $6.6^\circ$ & 0.2 \\
21:30-22:11 & 23.2 & 62.5 & $6.9^\circ$ & 0.1 \\
\cutinhead{Experiment 3}
17:01-17:56 & 1444.9 & 337.8 & $30.4^\circ$ & 0.1 \\
18:03-18:56 & 1511.1 & 213.5 & $26.8^\circ$ & 0.0 \\
19:11-21:11 & 1507.3 & 269.9 & $30.3^\circ$ & 0.0 \\
21:30-22:11 & 1498.8 & 169.9 & $31.4^\circ$ & 0.0 \\
\enddata
\tablecomments{The numbers are the {\it r.m.s.}differences between the ``correct'' magnetic fields described in section 4.2 and the ``incorrect'' magnetic fields. The calculations are carried over pixels having the $B_\parallel>10$ and $B_\bot>500$ G. Experiment 1: ``incorrect'' fields are inferred from time-averaged Stokes images without spectral line registration; Experiment 2: ``incorrect'' fields are inferred from Stokes spectral images with wavelength shifting 0.2 \AA~(5 wavelength positions); Experiment 3: ``incorrect'' fields are inferred from using the wrong Stokes $V$-signs.}
\end{deluxetable}

\subsection{Effects of the $V$-Sign Error}

The sign of the Stokes $V$ is set by the polarization modulator in the
IVM, but we find that the instrumental definition is unfortunately opposite to the conventional definition for data taken after January 1999. The standard representation of polarized light \citep{shurcliff1962} implies that Stokes $V$ is negative in the red
wing of an absorption line for a magnetic field pointing towards the
observer \citep{rees1987,rees1989}. This follows the analysis of
polarized radiative transfer \citep{unno1956} and is consistent with
laboratory observations of the Zeeman effect. From the MDI magnetogram \citep{1995SoPh..162..129S}, the NOAA10001 major sunspot polarity is
positive. Instead of being negative, Stokes $V$ was found positive in the
red wing (see the lower-right plot, ``Stokes V'', in the bottom panel
in Figure \ref{stokes_noise}).

If uncorrected in the analysis, the reversed sign of the IVM-$V$ leads to an over-correction of the magneto-optical ``Faraday'' effect. This is because Faraday rotation occurs when the left- and right-circular polarization propagate at different speeds, causing the plane of the linear polarization
to rotate. Such rotation in the linear polarization is proportional to the intensity of the line-of-sight magnetic field component, and is correlated with the sign of the component determined by the Stokes $V$. In a previous version of the IVM software (for instance, ``Triplet'' code), the sign correction was done on the longitudinal magnetic field $B_\parallel$ {\it after} the Stokes inversion, so that the sign of $B_\parallel$ as determined by IVM was consistent with results from other instruments. However, this is not the correct approach to the problem because it causes inaccurate transverse fields. 

The effects of the reversed Stokes $V$-sign is examined in the same way as for the effects of the spectral line shifts. The {\it r.m.s.}differences are calculated between the ``correct'' magnetograms and the ones inferred from original IVM Stokes $V$-sign. The results are summarized in Table \ref{unc} under ``Experiment 3''. The reversal of the $V$-sign relative to the conventional definition not only introduces abundant spurious field azimuths, but also changes the
strengths of both longitudinal and transverse components of the magnetic field.  The Stokes $V$-sign {\it must} be flipped before applying the magnetic field inversion code to the Stokes spectral images. 

\section{Vertical Electric Current Densities}

The goal of this work is to determine high-quality vector magnetic fields in the photosphere. In one application, 
we use the vector magnetic fields to calculate the vertical 
electric current densities for two active regions:  NOAA
10001, observed 20 June 2002 UT, was a unipolar
sunspot region with little change in appearance through the day; and NOAA 10030, observed 15 July 2002 UT,  was a large
quadrupolar sunspot complex showing constant flux convergence, 
emergence and cancellation during the course of the day.  NOAA 10030 produced an X3 flare accompanied by a white light flare and coronal mass ejections and is the subject of several research papers \citep{liu2003, gary2004, 2004ApJ...602..446G,2005ApJ...620.1092L, 2005A&A...438.1099H,2008ApJ...684..747T}.

\subsection{Absolute Vertical Electric Current Density}

In SI units, Amp\`{e}re's law reads $\mu_0 {\bf J} = \nabla \times
{\bf B}$, where ${\bf J}$ is the current density and $\mu_0$ is the
permittivity of the vacuum. The vertical component of ${\bf J}$
satisfies
\begin{equation}
\mu_0 J_z = (\nabla \times {\bf B})_z = \frac{\partial B_y}
{\partial x} - \frac{\partial B_x} {\partial y} ,
\label{eq:j}
\end{equation}
Since $\mu_0 = 4 \pi \times 10^{-7}$ [T m A$^{-1}$], we can write
$4 \pi J_z = (\nabla \times {\bf B})_z$, where the magnetic field
is measured in G and the current density is measured in mA m$^{-2}$, and the distance is measured in m.
Determining $J_z (x,y)$ on the photosphere requires knowledge of the
direction of the transverse field $(B_x,B_y)$, i.e., a solution of
the {\one} ambiguity problem. However, \citet{1998A&A...331..383S}
showed it is possible to calculate the absolute value of $J_z$
without solving the {\one} ambiguity problem. They derived the following
expression for $J_z^2 (x,y)$:
\begin{equation}
{(4\pi J_z})^2 = | (\nabla \times {\bf B})_z |^2 = B_x^2 g_y^2 +
B_y^2 g_x^2 - 2 (B_x B_y) g_x g_y ,
\end{equation}
where
\begin{eqnarray}
g_x & \equiv & \frac{1}{B_\bot^4} \left[ B_y^2 \frac{\partial (B_xB_y)}
{\partial y} - \frac{1}{2} B_xB_y \frac{\partial(B_y^2-B_x^2)}
{\partial y} - \frac{1}{2} B_\bot^2  \frac{\partial B_y^2} {\partial x}
\right ] ,\\
g_y & \equiv & \frac{1}{B_\bot^4} \left[ B_x^2 \frac{\partial (B_xB_y)}
{\partial x} - \frac{1}{2} B_xB_y \frac{\partial(B_x^2-B_y^2)}
{\partial x} - \frac{1}{2} B_\bot^2  \frac{\partial B_x^2} {\partial y}
\right] 
\end{eqnarray}

From the equations (\ref{eq:inclaz}) and (\ref{eq:mag}),  the observable quantities from the IVM are the vertical field component, $B_z$ (or $B_{\parallel}$); the transverse field strength $B_{\perp}$; and the magnetic azimuth, either $\phi$ or $\phi+180^{\circ}$. The two perpendicular horizontal components are written $B_x=B_\bot \cos \phi$ and $B_y=B_\bot \sin \phi$,  \one~ambiguity in the  $\phi$ is equivalent to the observable quantity $B_xB_y=\frac{1}{2}B_{\bot}^2\sin 2\phi$,  and $B_{\perp}=\sqrt{B_{x}^2+B_{y}^2}$.  Neither $B_x^2$ nor $B_y^2$ varies with the $\phi$  and $\phi+180^\circ$.
In this sense, the IVM provides all necessary quantities for calculating the absolute vertical electric current density, $|J_{z}|$, which can be determined without solving the ambiguity problem. Another condition is that the active region needs to be near the disk center so that the longitudinal fields approximate to the vertical fields, ${\bf B_z}$, and the transverse fields approximate to the horizontal fields, ${\bf B_\bot}$. 

\subsection{Absolute Vertical Current Density in NOAA 10001}
 
\begin{figure}[]
\begin{center}
\includegraphics[width=1.0\textwidth]{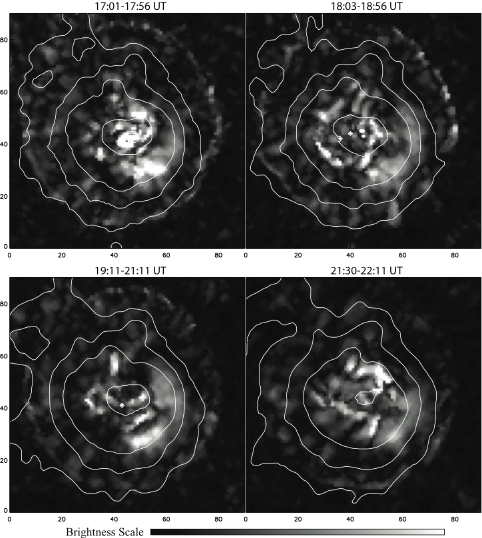}
\caption{The absolute electric current densities are displayed in the background. The superimposed contours represent the longitudinal magnetic field $B_\parallel=\pm 150,\pm 600, \pm 1200,\pm 2400$ G. The ``brightness scale'' bar in the bottom represents the $|J_z|$ ranging from 0 to 30 [mA m$^{-2}$]. The image sizes are $50\arcsec\times 50\arcsec$. \label{jz_ar10001}} 
\end{center} 
\end{figure}

Because the active region NOAA 10001 was near the disk center (N20W07) on 20 June 2002, the transverse fields are close to the horizontal magnetic fields and the longitudinal fields to the vertical magnetic fields. The absolute vertical current density maps are calculated from four vector magnetograms (see Fig. \ref{jz_ar10001}). The $|J_{z}|$ noise level is estimated in a similar way to the uncertainty of the magnetic azimuth for NOAA 10001. The pixels included in the estimation are those having transverse field strengths between 500-1500 G.  {\it r.m.s.} of $|J_z|$ differences are calculated between neighboring $|J_z|$ in time sequences, {\it i.e.}: $|J_z|$(17:01-17:56 UT)-$|J_z|$(18:03-18:56 UT),... $|J_z|$(19:11-21:11 UT)-$|J_z|$(21:30-22:11 UT), respectively. The average noise level is 6.7 mA m$^{-2}$ for NOAA 10001.

The strongest currents occur in localized regions at the
umbral-penumbral boundary where the longitudinal field is about
2000 G, and these currents vary significantly with time. More
persistent but weaker currents are found in the west and the south-west quadrant of the penumbra with magnitudes $\sim 15$ and $\sim 30$ mA m$^{-2}$. These $|J_{z}|$ values are well above the noise level. Therefore, we believe these signals represent real
electric currents on the Sun. On the other hand, bright rings surround the spots outside the penumbrae have $|J_{z}| \sim 10$ mA m$^{-2}$. They  are artifacts of the two phases used in the ``Triplet'' code, which are amplified by the derivatives approximating to finite differences in the $|J_z|$ calculation.

\subsection{Absolute Vertical Current Density in NOAA 10030}

\begin{figure}[]
\begin{center}
\includegraphics[width=0.95\textwidth]{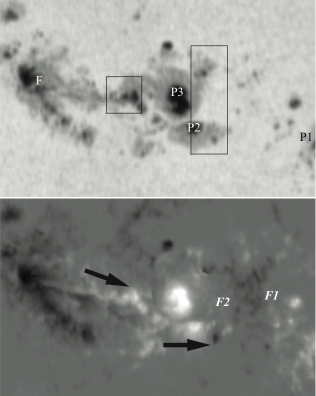}
\caption{IVM continuum image image (top) and the IVM longitudinal magnetic field Continuum images (bottom) for the NOAA 10030. Both images are taken at 17:35 UT on 15 July 2002 in the best seeing condition (1\arcsec).  In the top panel, P1, P2, P3 and F were three proceeding sunspots, and the following sunspot, respectively. Two boxes have the size 50\arcsec$\times$50\arcsec~and 50\arcsec$\times$180\arcsec, respectively. In the lower panel, {\it F1} and {\it F2} represent two filament locations residing above the magnetic field neutral lines. Black arrows indicate the magnetic field converging motion in the region. Image sizes are $240\arcsec \times 150\arcsec$.  \label{ar10030}} 
\end{center} 
\end{figure}

Figure \ref{ar10030} shows active region NOAA 10030 on 2002 July 15 UT. The spots are marked with ``P1'', ``P2'', ``P3'', and ``F'', for preceding and following sunspots, respectively, in the upper-left panel over the MDI continuum. During the course of the day, the major motions of the region showed the positive magnetic polarity converging in the directions shown by the black arrows in the upper-right panel of the figure. These motions were in close coincidence with the emerging magnetic flux, and the flux cancellation around the spot ``P3'' which also rotated counterclockwise. The flare started in the plage area between the filaments  {\it F1} and {\it F2}, and the filament {\it F2} erupted upon flare impulsive onset. 

Vector magnetograms for NOAA 10030 were produced using the same procedure as described for NOAA 10001. The
IVM raw data were first processed with the IVM standard software package. The ``seeing'' criterion, $\epsilon>0.0025$, is used for the AR10030, which selected $\sim 40$\% of the total observed data for further analysis. The Stokes $Q,U,V$ uncertainties are $2\times 10^{-3}$ for a single ``good seeing'' data set in the quiet Sun. The active region evolved rapidly with time. However, time-averaging Stokes images over the scale of an hour is still a good choice for increasing the signal-to-noise ratio while maintaining the recognizable fine structures at the same time. For all ``good seeing'' data, the IVM data are divided into three groups: 17:30-18:36 UT (29 data sets); 18:36-19:30 UT (18 data sets); and 20:30-22:00 (18 data sets). Of the three data sets, the first two were taken before a Class X3 flare at 20:03 UT and one of them after the flare. Data taken between 20:00-20:30 UT were not processed for magnetic fields because of line distortion by the flare. Nine data sets taken after 22:00 UT also satisfied the seeing criterion, but the Stokes spectral lines were badly distorted making it difficult for the magnetic field inversion. These data were also abandoned.  Three sets of time-averaged Stokes images were generated by registering Stokes images in both spatial and spectral dimensions within each group. The Stokes uncertainties are $4-5 \times 10^{-4}$. 

\begin{figure}[]
\begin{center}
\includegraphics[width=0.7\textwidth]{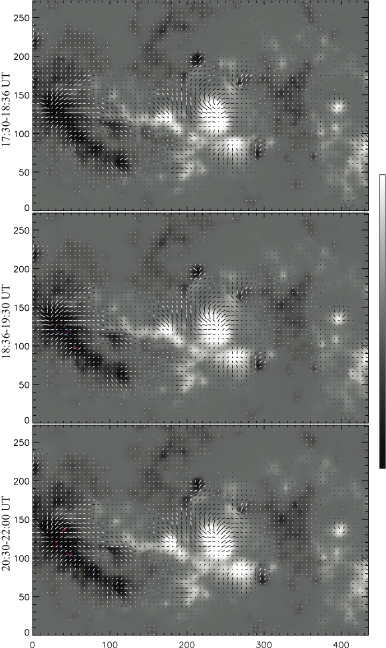}
\caption{NOAA 10030 vector magnetograms inferred from three time-averaged Stokes images. Backgrounds show the line-of-sight magnetic fields which strengths are shown in the brightness scale on the right, [-2000,2000] G. The transverse fields are represented by short bars: white bars against the negative longitudinal polarity, and the black bars against the positive polarities. Plotted pixel interval is 6 IVM pixels ($0.55 \arcsec \times 0.55 \arcsec$ per pixel) which corresponds to $B_\bot=1500$ G. Pixels having $B_\bot < 43$ G are not plotted. Image sizes are 240\arcsec $\times$ 150\arcsec. 
\label{ar10030-imgm}} 
\end{center} 
\end{figure}

Before the magnetic field inversion procedure, the Stokes $V$ signs were multiplied by  ``-'' sign. Locating pixels where the Stokes signals $<5\times 10^{-4}$, the average uncertainties of the magnetic field are 14 G for the longitudinal component, and 44 G for the transverse component. The azimuth uncertainty was estimated in the same way as that used for NOAA 10001, it is $7.5^{\circ}$. Fig.\ref{ar10030-imgm} shows the three final vector magnetograms from 3 time intervals. 

\begin{figure}[]
\begin{center}
\includegraphics[width=0.75\textwidth]{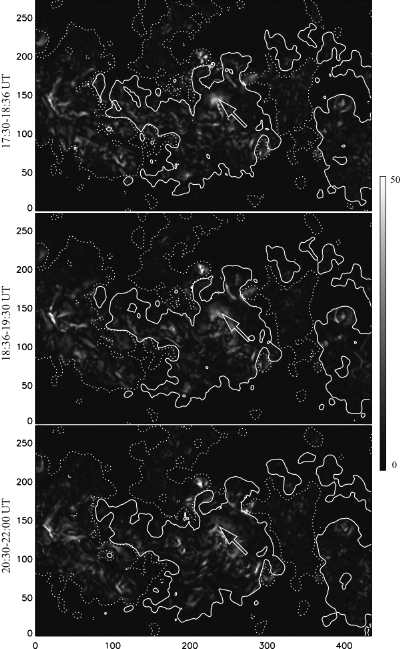}
\caption{The absolute vertical current densities are the background images with the NOAA 10030. The strengths are indicated by the vertical brightness scale on the right, 0 - 50 [mA m$^{-2}$]. The contours represent the line-of-sight magnetic field, [-100,100] G. Arrows indicate the same region having the $|J_{z}|=32, 24, 18$ mA m$^{-2}$ at different time intervals from the top to the bottom plots. Image sizes are 240\arcsec $\times$ 150\arcsec. \label{jz_ar10030}} 
\end{center} 
\end{figure}

NOAA 10030 was also near the disk center on 15 July 2002 (N19W01). The transverse fields approximate to the horizontal fields and the longitudinal fields approximate to the vertical fields. Three maps of the absolute value of the vertical current density, $|J_z|$, are shown in Figure \ref{jz_ar10030}. $|J_{z}|$ uncertainties are estimated in the same way as that estimated for the NOAA 10001. $|J_{z}|$ uncertainty is estimated as 7.0 mA m$^{-2}$ which is very similar to the uncertainty of NOAA 10001. As an example, we examine the northern area of the ``P3'' spot penumbra which spot rotated $\sim 29^{\circ}$ a day (see the white arrows in the figure \ref{jz_ar10030}) \citep{2005ApJ...620.1092L}. The area had average current densities $\sim 32$ mA m$^{-2}$ at 17:30-18:36 UT, $\sim 24$ mA m$^{-2}$ 18:36-19:30 UT, and dropped to $\sim 18$ mA m$^{-2}$ at 20:30-22:00 UT. These numbers are well above the current noise level, implying that the change was real. 

\subsection{Relations between Magnetic Field and Current Density}

\begin{figure}[]
\begin{center}
\includegraphics[width=1.0\textwidth]{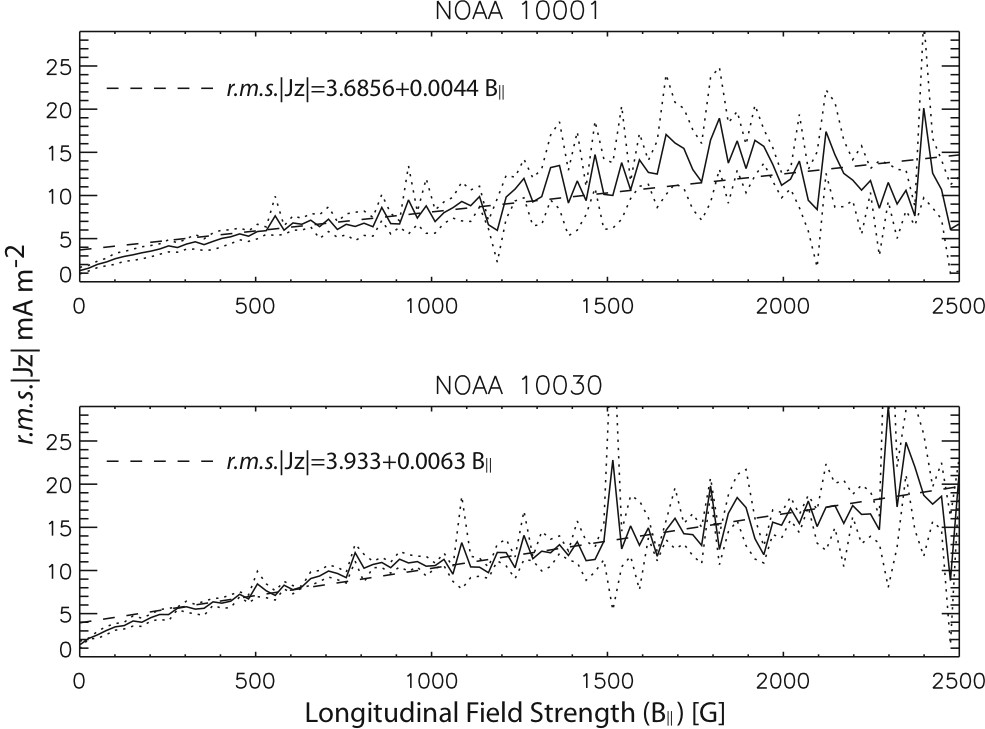}
\caption{{\it r.m.s.}($|J_z|$) are plotted as functions of the longitudinal magnetic field ($B_\parallel$) for NOAA 10001 on 20 June 2002 (top) and 10030 on 15 July 2002 (bottom). The {\it r.m.s.}($|J_z|$) are calculated over pixels having $B_{\parallel}\pm 5$ G. The solid curves represent the average {\it r.m.s.}($|J_z|$) among magnetograms of four time intervals 17:01-17: 56 UT, 18:03-18:56 UT, 19:11-21:11 and 21:30-22:11 UT for NOAA 10030; and magnetograms of three time intervals 17:30-18:36 UT, 18:36-19:30 UT, and 20:30-22:00 UT for NOAA 10030. The dotted curves are the average {\it r.m.s.}$|J_{z}| \pm \sigma$. The dashed curves are the linear least-square fits to the solid curves. 
\label{rmsjz-bl}} 
\end{center} 
\end{figure}

To investigate the correlation between the current density and the magnetic field, we examine the {\it r.m.s.}$|J_z|$ over pixels having $B_\parallel \pm \Delta B_\parallel$. Based on the uncertainty analyses,  $\Delta B_\parallel= 5$ G.  Each magnetogram corresponds to a curve {\it r.m.s.}$|J_z|$ varying with $B_\parallel$. Figure \ref{rmsjz-bl} shows relations for NOAA 10001 (top) and NOAA 10030 (bottom), respectively. Instead of curves of each magnetogram, the solid curves are the average {\it r.m.s.}$|J_z|$ among four magnetograms for NOAA 10001 (see Fig. \ref{imgms}), and three magnetograms for NOAA 10030 (see Fig. \ref{ar10030-imgm}); the dotted curves are the average {\it r.m.s.}$|J_z| \pm \sigma$ (standard deviation of {\it r.m.s.}$|J_z|$ among the magnetograms); the dashed lines are the linear least-square fits to the solid curves for {\it r.m.s.}$|J_z|$ above the uncertainty levels. The calculations are carried out over the same areas for both active regions, namely 240\arcsec$\times$150\arcsec. This area is slightly smaller than the images in Fig. \ref{fig1} for NOAA 10001, but is the same as the images in Fig. \ref{ar10030} for AR 10030.

Both active regions show linear correlations between $|J_z|$ and $B_\parallel$ in the form {\it r.m.s.}$|J_z|=a+b\times B_\parallel$, where $a=3.6856$  and $b=0.0044$ for NOAA 10001; $a=3.933$ and $b=0.0063$ for NOAA 10030, respectively.  Recall the magnetic field force-free field condition, $\nabla \times {\bf B} = \alpha {\bf B}$, where $\alpha$ is a scalar varying with space. This leads to the vertical component of the equation $4\pi J_z=\alpha B_z$, where $J_{z}$ is measured in mA m$^{-2}$, $B_{z} \approx B_\parallel$ in the current work measured in Gauss, and $\alpha$ is measured in m$^{-1}$. Allowing the force-free field condition valid, the figure \ref{rmsjz-bl} suggests that the vertical current density has a general relation with $B_\parallel$
\begin{equation}
|J_z| = J_{0}+\frac{\alpha}{4 \pi} |B_\parallel|
\label{eq:lff}
\end{equation}
where $J_{0}=a$, and $\alpha / 4\pi=b$.  $J_0 \neq 0$, but are less than vertical current uncertainties in both active regions. 

We should be cautious with how to interpret the equation \ref{eq:lff} from observations, which seems to suggest that the photospheric magnetic field is linear-force-free, which is an extreme case requiring a constant-$\alpha$ everywhere in the field. Meanwhile, it is still controversial whether the photospheric magnetic field is force-free \citep{1995ApJ...439..474M,2002ApJ...568..422M}. The figures \ref{jz_ar10001} and \ref{jz_ar10030} clearly show that the $\alpha$ is not constant because there are many fine structures in the vertical current density. To further verify non-constant-$\alpha$ within the active region, we isolate two areas in the AR 10030, which are marked with boxes on the top image in figure \ref{ar10030}.  The correlation parameters $b=-0.0020$ (the left box, 50\arcsec$\times$50\arcsec) and $b=0.0031$ (the right box, 50\arcsec$\times$180\arcsec), respectively. These local $b$s are very different from the general $b=0.0063$ derived from bigger region 240\arcsec$\times$150\arcsec. On the other hand,  the local $b=0.0040$ which is computed around the major sunspot of NOAA 10001 as those areas shown in figure \ref{imgms} (50\arcsec$\times$50\arcsec). This is very similar to the general $b$ derived with the bigger area 240\arcsec$\times$150\arcsec. 

Our observations show that $|J_z|$ increases with increasing $B_\parallel$ in the photosphere in both active regions in large scales, but NOAA 10030 is far from being modelled by a linear force-free field. In large scale, $|J_z|$ increases 1.4 times faster with increasing $B_\parallel$ in NOAA 10030 than in NOAA 10001. The ratio of rates of increase varies between 1 and 2 when the standard deviations are included in the linear least-square fitting. Recall that NOAA 10030 produced an X3 flare with emission in the continuum for 6 minutes, and was followed by two CMEs. But NOAA 10001 was a simple sunspot region without flare activity. It will be interesting to conduct a statistical study using a large sample of active regions.  The rate of increase may have the potential to forecast the flare productivity of active regions.  

The significant correlations between the total vertical currents and the total magnetic flux were reported in a statistical study of over 1000 active regions by \citet{leka2007}. However, they did not report any differences of correlation parameters among different regions.  
It could be that they used an overly simplistic magnetic field inversion-method - ``wavelet'' method.  We also note that the $\alpha_{AR}$ as the constant-$\alpha$ in the linear force-free field was sought by \citet{1999SoPh..188....3L} for a single active region, but no correlation was found between the $J_z$ and $B_\parallel$. They used the ``Derivative'' method to infer the vector magnetic field
It would seem prudent to conduct a new investigation of the relation between $B_\parallel$ and $|J_z|$ making full use of the current understanding of the Mees IVM. 

\section{Summary}

We present the procedures needed to obtain vector magnetic fields from
 the Imaging Vector Magnetograph at Mees Solar Observatory. The general
data reduction can be undertaken with the IVM software package
\citep{labonte1999} but extra steps are needed to deal with two issues with  IVM
and so to generate high quality Stokes images and vector magnetograms: 

(1). Large, previously undiscussed time-dependent spectral line shifts, probably due to instability of the
\FP~inside IVM, should be measured. The effects of the wavelength shift affect more the magnetic field strengths than the magnetic azimuth or the Doppler velocity.  

(2). A previously undiscussed sign-error built into the IVM definition of the Stokes $V$ parameter must be properly
corrected. Failure to take this step leads to an over-correction for Faraday rotation and to an over-estimated electric current density. In our work, the uncorrected $V$-sign error introduces average errors $\Delta B_\parallel\sim 1490$, $\Delta B_\bot \sim 248$ G, and $\Delta \phi \sim30^{\circ}$.  
Evidence suggests that the sign error occurred after January 1999 when IVM data have image sizes $512\times 512$. But a careful Stokes-$V$ sign verification using the longitudinal observations with other instruments is recommended. When Stokes-$V$ sign is found to be wrong, the Stokes-$V$ should be flipped by ``-'' first, then the magnetic fields are inferred from the Stokes images. 

(3) Further employing only data taken under good seeing conditions, and  correcting for the above two errors in the IVM data, 
we obtained formal uncertainties on the Stokes
$Q,U,V$ parameters for the quiet Sun of
1$\sigma \sim$  $1\times 10^{-3}$ for NOAA 10001 and 1$\sigma \sim$ $5\times 10^{-4}$ for NOAA 10030.  
The resulting uncertainties are 10 G for the
longitudinal magnetic field component, 40 G for the transverse component,
and $\sim 9^{\circ}$ for the magnetic azimuth in the sunspot penumbra.  


Absolute vertical electric current densities, $|J_{z}|$ [mA m$^{-2}$] are
calculated for the simple, flare-free spot NOAA 10001 on 20 June 2002 UT
and the complex active region NOAA 10030 on 15 July 2002 UT.  The calculation does not require disambiguation of the $180^{\circ}$ in the transverse field directions.

(1) The uncertainty on $|J_{z}|$ is $\sim$7.0 mA m$^{-2}$ for both NOAA 10001 and NOAA 10030.
 
(2) In large scale, the vertical current densities, {\it r.m.s.}($|J_{z}|$), increase with increasing longitudinal magnetic field strength ($B_{\parallel}\approx B_{z}$) in the form $|J_z|=a+b B_\parallel$.  The increasing rate  is $1- 2$ times larger in NOAA 10030 than in NOAA 10001. 

(3) Locally, the linear correlation parameters between $|J_z|$ and $B_\parallel$ are largely variable from place to place in complex active region (NOAA 10030), but are fairly constant in the simple region (NOAA 10001). 

\acknowledgments
Authors would like to thank the Mees observer, Garry Nitta, for his dedicated observing work. JL would like to thank David Jewitt for his critical comments throughout the paper, and the kind help with English editing. We thanks comments made by the referee, which helped to improve the paper. JL is supported by the NASA grant ``The Non-potential Structure of
Active Regions'' awarded to SAO which subcontracts to University of
Hawaii. She is also supported by the NASA grant  NNG06GE13G to the University of Hawaii to support the operation of the Mess Solar Observatory. The work of AvB is supported by NASA grant NNG05GK32G to
Smithsonian Astrophysical Observatory.

\appendix

\section{Magnetic Inversion Method for IVM Data}

Several methods have been developed to infer the vector magnetic
field on the Sun from polarization measurements of Zeeman-split
spectral lines (see section 1). The ``Triplet'' method is based on
a simple model for the transfer of polarized light in the solar
atmosphere \citep{unno1956}. The model parameters (e.g., magnetic
field strength and direction) are varied until the best fit to the
observed Stokes profiles is obtained \citep{auer1977}. Here, we use
the formulation of the method given by \citet{landolfi1982}
(hereafter LL82). We describe the implementation of the method
for IVM, which produces images in four Stokes parameters
($I,Q,U,V$) and 30 wavelengths in Fe~I 6302.5 {\AA}.
The implementation takes advantage of the fact that the
thermodynamic parameters of the model can be estimated directly from
the observed continuum intensity. The ``Triplet'' code was originally
developed by the late Barry J. LaBonte.

When a magnetic field is present, the atomic energy levels with
angular momentum quantum numbers $J > 0$ are split into multiple
components with magnetic quantum numbers $m = -J, \cdots, +J$.
LL82 described the formation of a ``normal'' Zeeman triplet in which
$J=0$ for the lower level (no Zeeman splitting) and $J=1$ for the
upper level. In the case of Fe~I 6302.5 {\AA}, the upper level has
$J=0$ and the lower level has $J=1$, but this reversal of the split
and unsplit levels compared to a normal triplet does not affect the
emergent Stokes profiles. Therefore, the formulae presented by LL82
can also be used for the 6302.5 {\AA} line.

The line formation model used by LL82 makes several assumptions.
First, the emitting atoms are assumed to be in Local Thermodynamic
Equilibrium (LTE), i.e., the populations of the atomic levels are in
agreement with the Boltzmann equation and the source function of the
emitted radiation is given by the Planck function, $B_\lambda (T)$,
where $T$ is the local temperature. Also, the solar atmosphere is
assumed to have plane-parallel stratification in the region of
emission, and the temperature $T(h)$ is assumed to decrease with
height $h$ such that the Planck function $B(\tau)$ at $\lambda = 6302$
{\AA} is a {\it linear} function of continuum optical depth, $B(\tau)
= B_0 + B_1 \tau$ (Milne-Eddington approximation). Here $B_0$ is the
value of the Planck function at the top of the photosphere
(temperature minimum region), and $B_1$ is the gradient of the Planck
function. Other parameters of the model are the magnetic field vector
${\bf B}$, Doppler shift $\Delta \lambda_0$ (due to mass flows on the
Sun), Doppler width $\Delta \lambda_D$, line broadening parameter
$\Gamma$, and certain ratios of line- and continuum opacity (see
below). These quantities are assumed to be independent of continuum
optical depth $\tau$. The assumption of LTE is reasonable in the
photosphere where collisional excitation of the atoms dominates over
radiative processes. The other assumptions are questionable because
observations show that the solar photosphere is inhomogeneous and
dynamic.

With the above approximations, the equations of radiative transfer
of polarized light in a Zeeman-split spectral line can be solved
analytically (LL82). The Stokes parameters of the radiation
emerging from a magnetic region on the Sun are given by
\begin{eqnarray}
I_{\rm p} (\Delta \lambda) & = & B_0 \left\{ 1 + \beta_0 \Delta^{-1}
(1+\eta_I)[ (1+\eta_I)^2+\rho^2 ] \right\} , \label{eq:Im} \\ 
Q_{\rm p} (\Delta \lambda) & = & - B_0 \beta_0 \Delta^{-1}
[ (1+\eta_I)^2 \eta_Q + (1+\eta_I) (\eta_V \rho_U - \eta_U \rho_V )
+ \rho_Q \chi ] , \label{eq:Qm} \\ 
U_{\rm p} (\Delta \lambda) & = & - B_0 \beta_0 \Delta^{-1}
[ (1+\eta_I)^2 \eta_U + (1+\eta_I) (\eta_Q \rho_V - \eta_V \rho_Q )
+ \rho_U \chi ] , \label{eq:Um} \\
V_{\rm p} (\Delta \lambda) & = & - B_1 \beta_0 \Delta^{-1}
[ (1+\eta_I)^2 \eta_V + \rho_V \chi ] , \label{eq:Vm} 
\end{eqnarray}
where subscript ``p'' refers to the polarized component of the
emission; $\Delta \lambda \equiv \lambda - \lambda_0$ is the
wavelength offset relative to the rest wavelength $\lambda_0$;
$\beta_0 \equiv \mu B_1 / B_0$; $\mu \equiv \cos \theta$ describes the
direction of propagation of the light relative to the radially outward
direction on the Sun; the $\eta_{I,Q,U,V}$ are ratios of line- and
continuum opacity; the $\rho_{Q,U,V}$ describe magneto-optical
effects; and the quantities $\Delta$, $\eta$, $\rho$ and $\chi$ are
defined by
\begin{eqnarray}
\Delta & = & (1+\eta_I)^2 [(1+\eta_I)^2 - \eta^2 + \rho^2 ]
- \chi^2 , \\
\eta^2 & = & \eta_Q^2+\eta_U^2+\eta_V^2 , \\
\rho^2 & = & \rho_Q^2+\rho_U^2+\rho_V^2 , \\
\chi & \equiv & \eta_Q \rho_Q + \eta_U \rho_U + \eta_V \rho_V .
\end{eqnarray}
The $\eta$'s and $\rho$'s are given by equation (2) and (3) of LL82:
\begin{eqnarray}
\eta_I & = & \onehalf \left[ \eta_p \sin^2 \Psi + \onehalf
(\eta_r+\eta_b)(1+\cos^2\Psi) \right] , \\
\eta_Q & = & \onehalf \left[ \eta_p - \onehalf (\eta_r+\eta_b)
\right] \sin^2 \Psi \cos 2\phi , \\
\eta_U & = & \onehalf \left[ \eta_p - \onehalf (\eta_r+\eta_b)
\right] \sin^2 \Psi \sin 2\phi , \\
\eta_V & = & \onehalf (\eta_r - \eta_b) \cos \Psi , \\
\rho_Q & = & \onehalf \left[ \rho_p - \onehalf (\rho_r+\rho_b)
\right] \sin^2 \Psi \cos 2\phi , \\
\rho_U & = & \onehalf \left[ \rho_p - \onehalf (\rho_r+\rho_b)
\right] \sin^2 \Psi \sin 2\phi , \\
\rho_V & = & \onehalf (\rho_r - \rho_b) \cos \Psi ,
\end{eqnarray}
where
\begin{eqnarray}
\eta_p & = &   \eta_0 H(a,v-v_0) , ~~~~
\eta_{b,r} =   \eta_0 H(a,v-v_0 \pm v_H), \\
\rho_p & = & 2 \eta_0 F(a,v-v_0) , ~~~~
\rho_{b,r} = 2 \eta_0 F(a,v-v_0 \pm v_H) .
\end{eqnarray}
Here $v \equiv \Delta \lambda / \Delta \lambda_D$, $v_0 \equiv
\Delta \lambda_0 / \Delta \lambda_D$ and $v_H \equiv \Delta
\lambda_H / \Delta \lambda_D$ are the wavelength offset, Doppler
shift and Zeeman splitting in units of the Doppler width; $a \equiv
\Gamma / (4 \pi \Delta \nu_D)$ is the damping constant; $\Gamma$ is
the line broadening parameter in frequency units; $\Delta \nu_D
\equiv c \Delta \lambda_D / \lambda_0^2$ is the Doppler width in
frequency units ($c$ is the speed of light); $H(a,v)$ and
$F(a,v)$ are the Voigt and Faraday-Voigt functions; and $\eta_0$ is
the ratio of line- to continuum opacity in the absence of a magnetic
field. The Zeeman splitting is given by $\Delta \lambda_H = 4.6686
\times 10^{-13} g_L \lambda_0^2 B$, where $B$ is the magnetic field
strength, $\lambda_0 = 6302.5$ {\AA}, and $g_L = 2.5$ is the Land\'{e}
factor for this line. The angles $\Psi$ and $\phi$ describe the
inclination and azimuth of the (constant) magnetic field relative to
the line of sight.

The observed Stokes profiles generally include a contribution from
instrumental stray light. Also, the magnetic field outside sunspots
is structured on subarcsecond scales that are not resolved by the IVM
instrument. Therefore, the observed profiles contain both polarized
and unpolarized components:
\begin{eqnarray}
I (\Delta \lambda) & = & f I_{\rm p} (\Delta \lambda) +
(1-f) I_{\rm s} (\Delta \lambda) , \label{eq:I} \\
Q (\Delta \lambda) & = & f Q_{\rm p} (\Delta \lambda) , ~~~
U (\Delta \lambda) = f U_{\rm p} (\Delta \lambda) , ~~~
V (\Delta \lambda) = f V_{\rm p} (\Delta \lambda) ,
\label{eq:QUV}
\end{eqnarray}
where subscript ``s'' refers to the unpolarized component of the
emission (straylight and/or non-magnetic contribution); $f$ is the
filling factor, i.e., the fraction of radiation due to the polarized
component; and $I_{\rm s} (\Delta \lambda)$ is the intensity profile
of the unpolarized component \citep{jls1989}. The latter is assumed
to be given by equation (\ref{eq:Im}) without a magnetic field:
\begin{equation}
I_{\rm s} (\Delta \lambda) = B_{0,s} \left[ 1 + \frac{\beta_0}
{1 + \eta_{0,s} H(a,v-v_{0,s})} \right] , \label{eq:Is}
\end{equation}
where $B_{0,s}$, $\eta_{0,s}$ and $\Delta \lambda_{0,s}$ are the
source function, line-to-continuum opacity ratio, and Doppler shift
characterizing the unpolarization component, and $v_{0,s} \equiv
\Delta \lambda_{0,s} / \Delta \lambda_D$.  The polarized and
unpolarized lines are assumed to have the same Doppler width $\Delta
\lambda_D$, line broadening parameter $\Gamma$, and source function
gradient $\beta_0$. The filling factor $f$ in equation (\ref{eq:QUV})
varies from pixel to pixel, and cannot be directly
determined from the observed Stokes profiles. We treat it as a free
parameter of the model.  Equations (\ref{eq:I}) and (\ref{eq:QUV})
give the Stokes profile prior to convolution with the instrumental
profile. To compare with observations we must convolve this profile
with the instrumental profile $P_n ( \Delta \lambda)$, where index $n$
indicates the wavelength setting of the IVM instrument ($n = 1,
\cdots, 30$).  The profile $P_n ( \Delta \lambda)$ is mostly
determined by the Fabry-Perot filter. Hence, the predicted Stokes
vector is
\begin{equation}
{\bf S}_n = \int {\bf S} (\Delta\lambda) P_n (\Delta\lambda) ~
d \Delta\lambda , \label{eq:Sn}
\end{equation}
where ${\bf S} (\Delta\lambda) \equiv [I,Q,U,V]$ is the Stokes profile
given by equations (\ref{eq:I}) and (\ref{eq:QUV}).

The above equations contain 10 unknown parameters ($B_0$, $\beta_0$,
$\eta_0$, $\Delta \lambda_D$, $\Delta \lambda_0$, $\Gamma$, $B$,
$\Psi$, $\phi$ and $f$). To determine these parameters, the
``Triplet'' code uses two different methods, one for plage regions
where the magnetic field strength $B$ and filling factor $f$ cannot
be independently determined (method I), and another for sunspots
where a non-linear least-square fitting procedure can be used
(method II). Method I is also used to obtain initial values of the
model parameters inside sunspots, as a starting point for non-linear
least-square fitting. Therefore, we describe this method first.

{\it Method I}: First, certain estimates of the thermodynamic
parameters are obtained for all pixels in the IVM image. Specifically,
the Doppler width is approximated by $\Delta \lambda_D \approx
(\lambda_0 / c) \sqrt{2kT/m + v_t^2}$, where $T$ is the brightness
temperature associated with the observed continuum intensity; $m$ is
the atomic mass of iron; and $v_t$ is a micro-turbulent velocity.
The latter is approximated by $v_t \approx [(T/2000)-1.3]$ km/s.
The line broadening is assumed to be dominated by van der Waals
broadening, $\Gamma = 8.08 C_6^{2/5} u_H^{3/5} n_H$, where $u_H =
\sqrt{2kT/m_H}$ is the most probable speed of the hydrogen atoms,
$n_H \approx 2 \times 10^{17}$ $\rm cm^{-3}$ is the total hydrogen
density at the height where the continuum is formed, and
$C_6 = 8 \times 10^{-32}$ \citep{1978stat.book.....M}.
The parameters $B_0$ and $\eta_0$ are estimated as
\begin{equation}
B_0 \approx \frac{I_c} {1 + \beta_0} , ~~~~
\eta_0 \approx \frac{1}{H(a,0)} \left( \frac{B_0 \beta_0}
{I_0 - B_0} - 1 \right) ,
\end{equation}
where $I_c$ is the observed continuum intensity, $I_0$ is the observed
line center intensity, and $\beta_0 = 1.3$. Next, to obtain more
accurate values of the thermodynamic parameters, a quiet area in the
IVM field of view is selected, and an expression similar to equation
(\ref{eq:Is}) is fit to the observed intensity $I_{\rm quiet} ( \Delta
\lambda )$ from the quiet region.  This yields fitted values of the
parameters $B_0$, $\beta_0$, $\eta_0$ and $\Delta \lambda_D$ for each
``quiet'' pixel. Then the mean of the estimated values of $B_0$ for
the ``quiet'' pixels is divided by the mean of the fitted values to
obtain a correction factor $C(B_0)$, and similar for the parameters
$\beta_0$, $\eta_0$ and $\Delta \lambda_D$.  Finally, the correction
factors are applied to the estimated values of $B_0$, $\beta_0$,
$\eta_0$ and $\Delta \lambda_D$ for all pixels, including plage and
sunspot regions ($\Gamma$ is unchanged from its estimated value). This
yields the final values of the thermodynamic parameters for method I.

We now describe how the magnetic parameters are determined in method
I. In facular areas outside sunspots, the Zeeman splitting is less
than the Doppler width and the Stokes profiles do not contain
sufficient information to independently determine both the magnetic
field strength ($B$) and the filling factor ($f$); only their product
$f B$ is well-constrained by the observations. Therefore, we assume a
fixed field strength of the magnetic elements, $B \approx 1500$ G.
Furthermore, we assume that the Stokes profile can be approximated as
follows:
\begin{equation}
Q (\Delta \lambda) = \tilde{q} \frac{Q_{\rm ref}(\Delta \lambda)}
{\sin^2 \Psi_{\rm ref}} , ~~~~
U (\Delta \lambda) = \tilde{u} \frac{Q_{\rm ref}(\Delta \lambda)}
{\sin^2 \Psi_{\rm ref}} , ~~~~
V (\Delta \lambda) = \tilde{v} \frac{V_{\rm ref}(\Delta \lambda)}
{\cos \Psi_{\rm ref}} ,
\label{eq:QUVm1}
\end{equation}
where $Q_{\rm ref} (\Delta \lambda)$ and $V_{\rm ref} (\Delta
\lambda)$ are reference profiles computed with equations (\ref{eq:Qm})
and (\ref{eq:Vm}) for some nonzero inclination angle $\Psi_{\rm ref}$
and azimuth angle $\phi_{\rm ref} = 0$. The parameters $\tilde{q}$,
$\tilde{u}$ and $\tilde{v}$ are defined by
\begin{equation}
\tilde{q} \equiv f ~ \sin^2 \Psi \cos 2\phi , ~~~~
\tilde{u} \equiv f ~ \sin^2 \Psi \sin 2\phi , ~~~~
\tilde{v} \equiv f ~ \cos \Psi , \label{eq:quv}
\end{equation}
where $\Psi$ and $\phi$ are the actual inclination and azimuth angles
on the Sun. Equations (\ref{eq:QUVm1}) are fitted to the observed Stokes
profiles using linear least-square fitting, which determines the
parameters $\tilde{q}$, $\tilde{u}$ and $\tilde{v}$. Then equations
(\ref{eq:quv}) are inverted as follows:
\begin{eqnarray}
t & \equiv & \sqrt{\tilde{q}^2+\tilde{u}^2} = f \sin^2 \Psi , ~~~~
s \equiv \sqrt{\tilde{q}^2+\tilde{u}^2+4\tilde{v}^2} =
f ( 2 - \sin^2 \Psi ) , \\
f & = & \onehalf (t+s), ~~~~ \Psi = \arccos (\tilde{v}/f) , ~~~~
\phi = \onehalf \arctan (\tilde{u}/\tilde{q}) .
\label{eq:inclaz}
\end{eqnarray}

{\it Method II}: This inversion method is used in sunspots. The spot
areas are identified by the differences of the continuum brightness
relative to the quiet Sun. For each pixel inside a sunspot, the
thermodynamic and magnetic parameters are determined by non-linear
least-square fitting of equation (\ref{eq:Sn}) to the observed
normalized Stokes parameters, $[Q/I,U/I,V/I]_n$ with $n = 1, \cdots,
30$. The fitting uses the Levenberg-Marquart algorithm \citep{press1994}. The observed intensities $I_n$ are not used in the
fit. Only certain parameters are allowed to vary in this fitting
process ($B_0$, $\eta_0$, $\Delta \lambda_0$, $B$, $\Psi$, $\phi$ and
$f$), while others are fixed to the values determined with method I
($\beta_0$, $\Delta \lambda_D$ and $\Gamma$). The reason is that the
observed Stokes profiles do not provide strong constraints on the
latter set of model parameters. The initial values of the variable parameters are those obtained from
method I. However, to ensure fitting convergence the initial filling
factors and field strengths are adjusted such that $f \le 0.8$.

Certain checks are carried out to make sure that the magnetic field
strength $B > 0$, the inclination angle $\Psi$ is in the range
$[0,\pi]$, and the filling factor is in the range $f \le 1$. For
pixels with parameter values outside these ranges the non-linear
fitting process may have failed to converge. The parameters at these
``bad pixels'' are assigned median values from surrounding pixels,
and the fitting is run again to obtain improved values. The azimuth
angle is adjusted such that $-\pi/2 < \phi < \pi/2$ i.e., no ambiguity
solution is performed.

The final output from the ``Triplet'' code (for both methods I and II)
gives the longitudinal and transverse fields, which are defined by
\begin{equation}
B_{\parallel} \equiv f B \cos \Psi , ~~~~
B_{\perp} \equiv f B \sin \Psi .
\label{eq:mag}
\end{equation}
Near solar disk center, $B_{\parallel}$ is essentially the magnetic
flux density in an IVM pixel. However, $B_{\perp}$ does not have
a clear physical interpretation. To obtain the transverse magnetic
field $B_\perp$ in the unresolved magnetic elements, one should divide
$B_{\perp}$ by the filling factor.

\end{document}